# A generalized ride-matching approach for sustainable shared mobility

Seyed Mehdi Meshkani[a,*], Bilal Farooq[a]

[a]*Laboratory of Innovations in Transportation (LiTrans), Ryerson University, Canada*

**Abstract**

On-demand shared mobility is a promising and sustainable transportation approach that can mitigate vehicle externalities, such as traffic congestion and emission. On-demand shared mobility systems require matching of one (one-to-one) or multiple riders (many-to-one) to a vehicle based on real-time information. We propose a novel Graph-based Many-to-One ride-Matching (GMOMatch) algorithm for the dynamic many-to-one matching problem in the presence of traffic congestion. GMOMatch, which is an iterative two-step method, provides high service quality and is efficient in terms of computational complexity. It starts with a one-to-one matching in Step 1 and is followed by solving a maximum weight matching problem in Step 2 to combine the travel requests. To evaluate the performance, it is compared with a ride-matching algorithm developed by Simonetto et al. (2019). Both algorithms are implemented in a micro-traffic simulator to assess their performance and their impact on traffic congestion in Downtown,Toronto road network. In comparison to the Simonetto, GMOMatch improved the service rate, vehicle kilometer traveled and traffic travel time by 32%, 16.07%, and 4%, respectively. The sensitivity analysis indicated that utilizing vehicles with a capacity of 10 can achieve 25% service rate improvement compared to a capacity of 4.

## 1. Introduction

According to the United Nations (UN), 68% of the world's population will live in urban areas by 2050 (Ritchie, 2018). Such rapid growth in urbanization increases the demand for trans-

*Corresponding Author.

*Email addresses:* smeshkani@ryerson.ca (Seyed Mehdi Meshkani), bilal.farooq@ryerson.ca (Bilal Farooq)

portation (Tafreshian and Masoud, 2020). To simultaneously satisfy this demand and alleviate the negative impacts of transportation (e.g, road congestion and emissions), more sustainable forms of transportation need to be conceived (Najmi et al., 2017; Chehri and Mouftah, 2019; Sachan et al., 2020).

Over the past few years, with the advancements in information and communication technology, emergence of smartphones, and ubiquity of high-speed internet, on-demand shared mobility services such as ridehailing and ridesharing have gained attention and have shown considerable growth (Agatz et al., 2011; Feng et al., 2017). Benefits of such modes include a point-to-point high level of service, decrease in traffic congestion and emissions, decline in parking space demand, and reduction in travel cost (Mourad et al., 2019; Najmi et al., 2017). However, some studies (Castiglione et al., 2018; Qian et al., 2020) showed that on-demand transportation companies such as Uber and Lyft contribute significantly to the increase in traffic congestion as well as vehicle miles traveled. To obtain sustainable benefits from such on-demand services, well-designed systems need to be developed that can utilize the supply optimally, while providing a high level of service.

The ride-matching problem as the core of ridehailing and ridesharing systems is the generalization of the dial-a-ride problem (DARP) that finds the best vehicle from a large pool of vehicles for a ride request (Yu et al., 2019; Tafreshian et al., 2020). Many-to-one dynamic ride-matching is needed in the on-demand shared mobility services, where one vehicle can serve multiple riders simultaneously. According to Lokhandwala and Cai (2018), with an increase in the vehicle occupancy rate due to ridesharing, the current taxi fleet can be reduced by 59% in NYC, while maintaining the service quality. This reduction can positively affect the urban areas from different aspects. Utilizing less taxi fleet in the cities can lead to the decrease in the traffic congestion as well as the vehicles emissions. Furthermore, individual’s transportation cost decreases and more travellers are encouraged to share their rides. Decrease in the car ownership rate and parking space demand are some other good impacts. Therefore, given all these benefits, developing a well-designed many-to-one ride-matching service can provide cities with sustainability from different environmental, social, and economical aspects.

Various studies have addressed the many-to-one ride-matching problem (Masoud and Jayakr-ishnan, 2017b; Alonso-Mora et al., 2017; Simonetto et al., 2019; Tafreshian and Masoud, 2020). Computational complexity and service quality are two important aspects of such problems. Masoud and Jayakrishnan (2017b) utilized the first-come, first-serve (FCFS) strategy to match riders with the vehicles. Although FCFS is efficient in terms of computational complexity, there is no guarantee to provide a ride-matching system with a high-quality level of service, especially in congested networks. Alonso-Mora et al. (2017) developed a multi-step algorithm for the many-to-one ride-matching problem, which provided high service quality. However, according to Simonetto et al. (2019), the proposed algorithm used a heuristic method to solve an integer optimization problem that may face scalability issues as its computational complexity is dependent on the vehicle capacity. Simonetto et al. (2019) proposed a ride-matching method based on the linear assignment problem. The method is computationally efficient. However, in each iteration, only a one-to-one matching problem is solved, and riders are not combined. In such a strategy, the quality of the service can not always be guaranteed. Tafreshian and Masoud (2020) proposed a graph partitioning method to solve the one-to-one ridesharing problem and then they applied the method on many-to-one ridesharing where one driver can serve several passengers. The algorithm was only demonstrated on low capacity vehicles (3 passengers). However, even with utilizing low capacity vehicles, no significant improvement in the performance was reported when compared to the one-to-one ridesharing. The resulting high number of clusters in such a large-scale problem adversely affected the optimality gap.

In order to resolve the issues in the aforementioned studies, in this study, we propose a novel graph-based heuristic algorithm for the dynamic many-to-one ride-matching problem that reduces the computational complexity of the problem, while providing a high-quality service in a congested network. The proposed Graph-based Many-to-One ride-Matching (GMOMatch) algorithm is iterative and consists of two steps. In the first step, riders are assigned to vehicles by solving a one-to-one ride-matching problem. In the second step, a maximum weight matching problem is solved to combine riders with similar itineraries through the idea of matching vehicles. It is assumed that all riders are willing to share their rides. Moreover, each ride request is considered a single passenger providing the ride-matching system with trip information. The proposed algorithm is implemented on an agent-based micro-traffic simulator to measure different indicators and examine the impact of ride-matching algorithm on traffic congestion. Finally, to evaluate our algorithm's performance, a comparison is conducted with the matching algorithm developed by Simonetto et al. (2019).

In brief, the major contributions of this paper include the following:

- A novel graph-based heuristic algorithm, *GMOMatch*, for the dynamic many-to-one ride-matching problem that lowers the computational complexity, while providing a high quality service. The proposed algorithm is flexible enough to accommodate multiple objective concerning users and service providers.
- A systematic theoretical complexity analysis is developed for GMOMatch and compared with the existing state-of-the-art, Simonetto's algorithm.
- Unlike most of the previous studies, the proposed algorithm is implemented in an agent based micro-traffic simulator to examine how the ride-matching algorithm affects traffic congestion in the presence of other vehicles.
- Extensive comparison of GMOMatch is developed with the Simonetto's algorithm in terms of service quality, sustainability, and computational complexity and detailed sensitivity analysis is undertaken over various variables and algorithm parameters.

The GMOMatch ride-matching algorithm can be used in different types of shared mobility services (static/dynamic) such as ridehailing, ridesharing, microtransit, and shuttles. In this study, we address its application on a dynamic ridehailing system.

The rest of this paper is organized as follows. In Section 2, we briefly review the relevant literature on ridehailing and ridesharing services. Section 3 introduces the dynamic ride-matching system, including system setting, system framework, and details of the ride-matching algorithm (GMOMatch). Section 4 presents the description of the case study, results and discussions. Finally, Section 5 concludes our findings and provides some directions for future research.

## 2. Background

Dynamic ridehailing and ridesharing are the two common types of on-demand shared mobility services. Dynamic ridehailing is a transportation service for compensation in which drivers and passengers are matched in real-time (Shaheen et al., 2019). In such systems, drivers unlike passengers do not have a tight time window (Tafreshian et al., 2020), which makes it more similar to a taxi service. Dynamic ridesharing is a service that connects drivers and passengers with similar itineraries and time schedules in order to split travel costs (Agatz et al., 2011; Shaheen et al., 2019).

According to Wang and Yang (2019), matching and order dispatching algorithms have a significant impact on the overall performance and efficiency of ridehailing systems. A well-designed matching algorithm provides a high-quality service for passengers and deploys vehicles more efficiently.

Ridehailing and ridesharing systems in the literature can be characterized by various features such as matching type, assignment approach, and modelling scale (Agatz et al., 2012; Tafreshian et al., 2020; Guériau et al., 2020). Some research efforts in the literature focused on the one-to-one matching problem in which a single driver/ vehicle can be matched with at most one rider (Agatz et al., 2011; Nourinejad and Roorda, 2016; Najmi et al., 2017; Lyu et al., 2019; Bertsimas et al., 2019; Özkan and Ward, 2020). Other studies addressed the many-to-one matching problem where a single vehicle/driver serves multiple riders (Jung et al., 2016; Alonso-Mora et al., 2017; Masoud and Jayakrishnan, 2017a,b; Qian et al., 2017; Simonetto et al., 2019). Assignment approach is another important feature that can be taken into account. Most of the studies in the context of ridesharing and ridehailing took a centralized assignment approach, where the ride-matching system is operated by a central entity responsible for all the requests. (Agatz et al., 2011; Liu et al., 2015; Qian et al., 2017; Alonso-Mora et al., 2017; Li et al., 2018; Bertsimas et al., 2019; Simonetto et al., 2019; Tafreshian et al., 2020). Other studies, however, deployed decentralized assignment approach in which the ride-matching system is operated locally by different entities or various types of decomposition and partitioning methods are utilized in order to convert the main ride-matching problem into several sub-problems that are computationally less expensive, while providing similar matching performance as the centralized approaches. (Nourinejad and Roorda, 2016; Sánchez et al., 2016; Najmi et al., 2017; Masoud and Jayakrishnan, 2017a; Baza et al., 2019; Kudva et al., 2020; Tafreshian and Masoud, 2020; Tafreshian et al., 2020).

Masoud and Jayakrishnan (2017a) proposed a decomposition algorithm to convert the original many-to-many ride-matching problem into smaller sub-problems that are tractable computationally. They introduced a pre-processing procedure to reduce the size of the optimization problem. Sub-problems are independent from each other which allows computations to be done in parallel. In another research effort, Masoud and Jayakrishnan (2017b) presented a real-time and optimal algorithm for many-to-many ride-matching problem, which aimed to maximize the number of served riders in the system. Their matching algorithm was based on First Come First Serve (FCFS), while the participant itineraries are determined using the dynamic programming. To improve the quality of the solution obtained by FCFS, they introduced a peer-to-peer exchange method. To assess the performance of the algorithm, they generated multiple random instances of ridesharing problem with different ratio of riders to drivers in a grid network. Qian et al. (2017) addressed the taxi group ride problem (TGR) to optimally grouping passengers with similar itineraries. They proposed three algorithms, including exact, heuristic, and greedy algorithms. To evaluate the performance of the algorithm, they used the taxi trajectory datasets from New York City (NYC), Wuhan, and Shenzhen. Their results showed that the heuristic outperformed the exact and greedy algorithms in terms of computational efficiency and solution quality.

Alonso-Mora et al. (2017) proposed a multi-step graph-based procedure to assign drivers to riders efficiently. Their algorithm allowed the use of low as well as high capacity vehicles. First, they created a shareability graph of requests and vehicles. Next, the graph of candidate trips, and vehicles that can execute them is created. Finally, using integer linear programming (ILP), they optimally assign requests to vehicles. The computational complexity of ILP is $O(mn^v)$, where $m$ is the number of vehicles, $n$ is the number of requests, and $v$ is the maximum capacity of the vehicles. They used NYC taxi data to evaluate the performance of their algorithm. Their results showed that 98% of the taxi rides could be served with just 3,000 taxis with a capacity of four, instead of the existing 13,000 taxis. Simonetto et al. (2019) used a federated architecture to linearly assign requests to vehicles. The proposed system consisted of a context-mapping algorithm to filter vehicles, a single dial-a-ride problem to obtain optimal travel route and associated cost, and a linear assignment problem to optimally match requests to vehicles. They used the NYC and Melbourne Metropolitan Area datasets to evaluate the system performance. They compared their algorithm with Alonso-Mora et al. (2017) and reported less computational complexity, while maintaining high-quality ride-matching level of service. Tafreshian and Masoud (2020) proposed a graph partitioning method for the one-to-one matching problem in bipartite graphs. It is then extended to a more complex case, where one vehicle can serve multiple riders (many-to-one matching). However, the computational complexity of such problem increased and an exhaustive evaluation was not practical. Thus, only two scenarios with a vehicle capacity of three were evaluated. Their results revealed that no significant improvement in the performance was achieved when compared to the one-to-one matching. Considering the high number of clusters in this NP-hard large size problem affected the optimality gap which was stated as one of the reasons for such a poor performance.

Another dimension of ridehailing and ridesharing systems in the literature is the modelling scale. Most of the studies in the context of ridehailing and ridesharing have been simulated and implemented at macroscopic/mesoscopic scale (Nourinejad and Roorda, 2016; Masoud and Jayakrishnan, 2017a; Alonso-Mora et al., 2017; Simonetto et al., 2019; Tafreshian and Masoud, 2020). One of the issues associated with these scales is that it is not clear how the proposed ride-matching system and traffic congestion affect each other in the presence of other vehicles (Guériau et al., 2020). Some recent studies in the context of shared automated vehicles (SAVs) used microsimulation to address how SAVs affects urban mobility systems (Dandl et al., 2017; Oh et al., 2020; Huang et al., 2020; Guériau et al., 2020). Dandl et al. (2017) deployed a microsimulation model to study how an autonomous taxi system affected the traffic network in the city of Munich. Oh et al. (2020) assessed the performance of shared driverless taxis including demand, supply and their interactions. On the supply side, they developed a heuristic matching and routing algorithm. Huang et al. (2020) explored the idea of using the SAVs to bring first-mile last-mile connectivity to transit in automated mobility districts. Guériau et al. (2020) developed a reinforcement learning-based shared autonomous mobility on-demand system for ridesharing and vehicle relocation where vehicles consider traffic congestion in their decisions. For the ride-matching, vehicles simply pick up the nearest passenger and no direct collaboration or coordination exist between vehicles.

Table 1 summarizes the major features of the recent literature on the ride-matching problem. As reviewed above, Tafreshian and Masoud (2020) utilized graph partitioning method to solve the one-to-one ride-matching problem and then extended it to the many-to-one ride-matching. Although, their algorithm performed well for the one-to-one matching, they reported unsatisfactory performance when it applied on the many-to-one ride-matching problem. Moreover, the study only addressed vehicles with the capacity of three, and medium and high capacity vehicles were not investigated. Guériau et al. (2020) developed a decentralized approach for the many-to-one ride-matching. However, their matching algorithm is greedy and vehicles are simply assigned to the nearest passenger. This method cannot guarantee to provide passengers with a high-quality service. Moreover, the study only considered vehicles with the capacity of four and the impact of vehicles with different capacities were not examined. Furthermore, they did not provide any computational complexity analysis or any comparison with a state-of-the-art algorithm. Our proposed algorithm, however, is in the class of algorithms that takes the centralized assignment approach. Alonso-Mora et al. (2017) and Simonetto et al. (2019) are some of the recent examples of such a class. Moreover, unlike Guériau et al. (2020) that deployed a simple greedy matching method, an advanced algorithm is proposed in this study that enhances the quality of the matching and consequently system performance. We also investigate the impact of various types of vehicles, including low, medium and high capacity. In Alonso-Mora et al. (2017), the computational complexity is dependent on the vehicle capacity, which may cause the scalability issue. The computational complexity of our proposed algorithm is fixed and is the same as that of the Simonetto algorithm by Simonetto et al. (2019). Nevertheless, unlike Simonetto et al. (2019), where each vehicle is matched with only one rider at each update time, our algorithm combines requests with similar itineraries, leading to utilizing the available vehicles more efficiently, improving the quality of service, and enhancing the traffic congestion. According to Table 1, most of the studies considered a macro/mesoscopic approach for the ride-matching problem. There is only one study Guériau et al. (2020) that utilized microscopic scale and addressed the many-to-one ride-matching problem which differs from this study from the aforementioned aspects.

In this paper, we introduce a novel many-to-one ride-matching algorithm for the dynamic on-demand shared mobility in the congested networks, which aims to improve the service quality, while considering the computational complexity. We deploy a micro-traffic simulator to examine the algorithm’s performance on the service quality and network traffic.

Table 1: Literature on ride-matching problem

| | Matching type | | Assignment | | Modeling scale | |
|---|---|---|---|---|---|---|
| **Study** | **one-to-one** | **many-to-one** | **centralized** | **decomposing /partitioning** | **macro/mesoscopic** | **microscopic** |
| Agatz et al., 2011 | * | | * | | * | |
| Nourinejad and Roorda, 2016 | * | | | * | * | |
| Najmi et al., 2017 | * | | | * | * | |
| Lyu et al., 2019 | * | | * | | * | |
| Bertsimas et al., 2019 | * | | * | | * | |
| Özkan and Ward, 2020 | * | | * | | * | |
| Jung et al., 2016 | | * | * | | * | |
| Alonso-Mora et al., 2017 | | * | * | | * | |
| Masoud and Jayakrishnan, 2017a | | * | | * | * | |
| Masoud and Jayakrishnan, 2017b | | * | * | | * | |
| Qian et al., 2017 | | * | * | | * | |
| Simonetto et al., 2019 | | * | * | | * | |
| Tafreshian and Masoud, 2020 | | * | | * | * | |
| Oh et al., 2020 | * | | * | | | * |
| Guériau et al., 2020 | | * | | * | | * |
| This study | | * | * | | | * |

## 3. Methodology

In the many-to-one ride-matching problem, one vehicle can serve multiple riders, while one rider can only be served by a single vehicle. This problem is a special case of general pick-up and delivery problem introduced by Savelsbergh and Sol (1995), and is known to be NP-hard (Qian et al., 2017; Tafreshian and Masoud, 2020). Therefore, for large-scale dynamic ride-matching problems, it is not practical to use exact methods to solve the problem in a short period of time. Thus, heuristic algorithms are proposed to solve this problem efficiently for large instances. In this section, we first introduce the general setup of a dynamic ride-matching system and its characteristics. We then propose the heuristic dynamic many-to-one ride-matching algorithm.

### *3.1. Dynamic ride-matching system setup*

We consider a set of ride requests $R = \{r_1, r_2, ..., r_m\}$ and a set of vehicles $V = \{v_1, v_2, ..., v_n\}$ with total capacity of *cap*. The ride-matching service aims to assign online ride requests to vehicles and find corresponding schedules while some constraints are satisfied. A ride request/rider refers to a person who places an order-mostly through a mobile application-to be picked up from an origin and to be dropped off at a destination. Furthermore, a passenger refers to a ride request that has been assigned to a vehicle. This passenger can be already on-board or can be waiting to be picked up. The set of passengers assigned to vehicle $v \in V$ is denoted by $P_v$. An available vehicle is a vehicle that has at least one empty seat. Each available vehicle $v \in V$ can be assigned no more than its current empty seats that is denoted by $cap_v^t$.

Each request $r \in R$ consists of a request-time $t_r$, origin $O_r$, and destination $D_r$. In addition, as in Agatz et al. (2011), we assume that each request $r$ provides an earliest departure time from their origin $e_r$ and latest time they would like to arrive at their destination $l_r$ (see Fig.1a). Furthermore, there is time flexibility $f_r$ which specifies the difference between rider's earliest departure time and the latest time he would like to depart $q_r = e_r + f_r$ and is computed as $f_r = l_r - e_r - H(O_r, D_r)$, where $H(O_r, D_r)$ is the travel time when rider directly (e.g. with privately owned vehicle) goes from his origin to destination. Without the loss of generality in order to make the system more dynamic, in this study, we assume that travel request time and their earliest departure time are the same $t_r = e_r$.

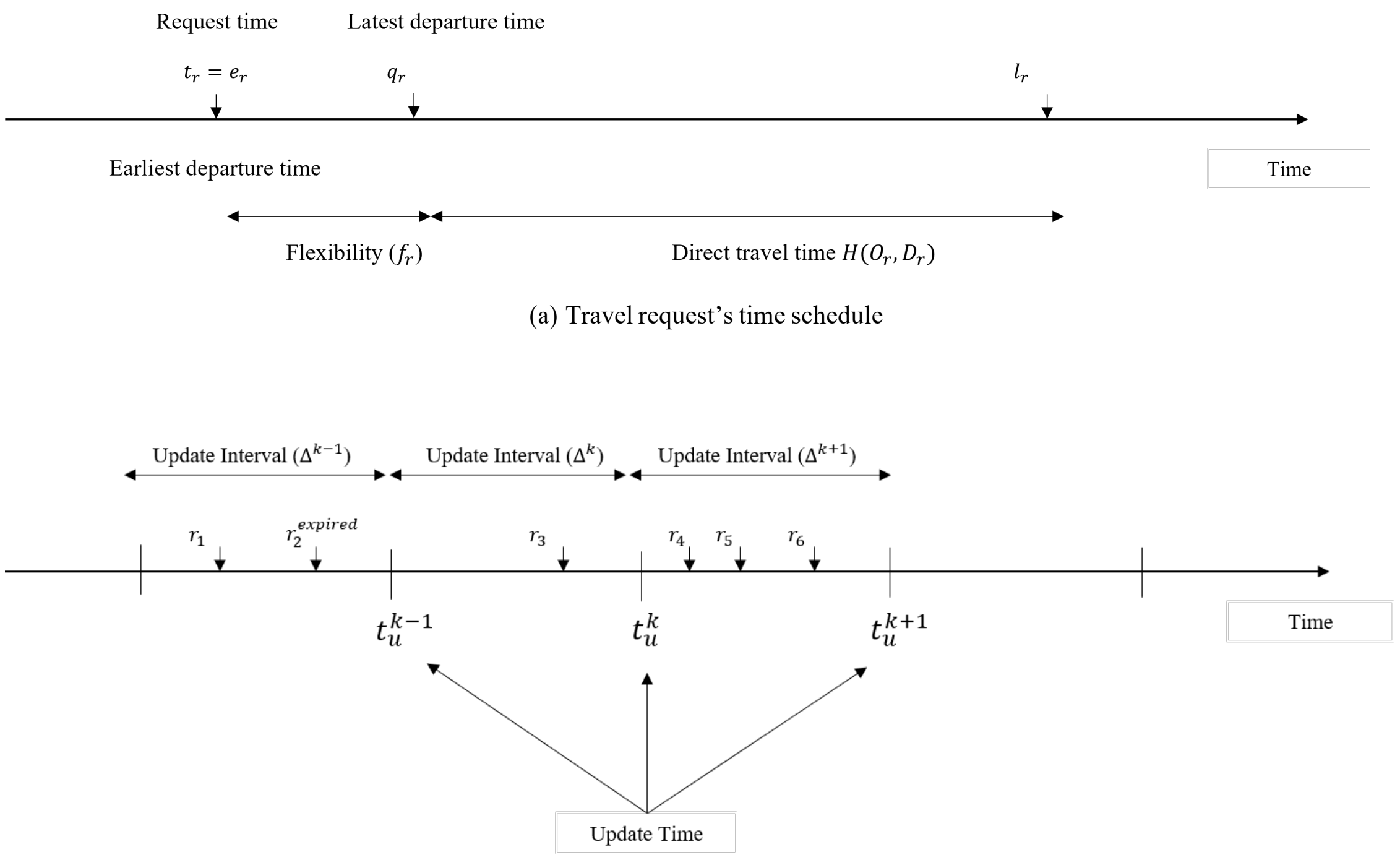

(a) Travel request's time schedule

(b) Rolling horizon structure

Figure 1: Ride-matching system setting

To solve the dynamic ride-matching problem in real time, we use the rolling horizon strategy suggested by Agatz et al. (2011). In this strategy, the ride-matching algorithm is solved periodically at specific time over fixed time intervals referred to as "update time" $t_u^k (k = 0, 1, 2, ...)$ and update interval $\Delta^k = t_u^k - t_u^{k-1}$ (see Fig. 1b). During each update interval, new riders place their orders to get a ride. At each update time, the system operator considers both new riders and those who have not been finalized or expired. A request is finalized when assigned to a vehicle and is expired when the current time $t$ exceeds the request's latest departure time $t > q_r$ while it has not been assigned to a vehicle yet. Rolling horizon iterations continue until all riders exit the system either by being matching or by having expired. Needless to say that current time $t$ is definitely greater or equal than riders' request time/earliest departure time $t \geq t_r = e_r$. As an example, in Fig.1b, when the system operator runs the ride-matching algorithm at current time $t_u^{k+1}$, it includes all of the new requests $\{r_4, r_5, r_6\}$ over update interval $\Delta^{k+1}$ and all requests $\{r_1, r_3\}$ related to previous update intervals $\Delta^k$ and $\Delta^{k-1}$ but $\{r_2\}$ which has been expired because the current time $t_u^{k+1}$ exceeds $r_2$'s latest departure time.

A set of constraints $Z$ consists of a capacity constraint $z_0$ and two time constraints $z_1$ and $z_2$. These constraints need to be satisfied so that one vehicle potentially is capable of serving a request. The first constraint, $z_0$, ensures that each vehicle has at least one empty space. In other words, it finds available vehicles. The second constraint, $z_1$, states that any request $r$ should be picked up not later than its latest departure time $q_r$ and the third constraint, $z_2$, expresses that the request $r$ should be dropped off not later than its latest arrival time $l_r$.

### *3.2. GMOMatch algorithm*

Given a set of ride requests $R$ and set of vehicles $V$ at current time $t$ from section 3.1, the many-to-one ride-matching problem is the matching of riders with vehicles such that each vehicle can be matched with multiple riders. In this problem vehicles can be low,medium, or high capacity. To solve this problem, we propose the GMOMatch which is a graph-based iterative algorithm that returns requests-vehicles matching and pick up/drop off scheduling. Before presenting the algorithm, two definitions are described.

**Definition 1:** Let $R_v$ be vehicle $v$'s assigned requests set. It is empty at the beginning of the matching at every update time $t_u^k$ $(\forall v \in V : R_v = \varnothing)$. Vehicle $v \in V$ is defined as an assigned vehicle if its assigned requests set is not empty ($R_v \neq \varnothing$). As an example in Fig. 2 (a) both vehicles $v_n$ and $v_n$ are assigned vehicles while in 2 (b) only $v_n$ is considered an assigned vehicle.

**Definition 2:** Two assigned vehicles $v_n$ and $v_n$ are matched with each other when assigned requests set of one of them depending on the direction of the link between them is allocated to the other one. As presented in Fig. 2, vehicle $v_n$ is matched with vehicle $v_n$ (a) which means that its assigned requests set is allocated to $v_n$ (b) and the requests are combined.

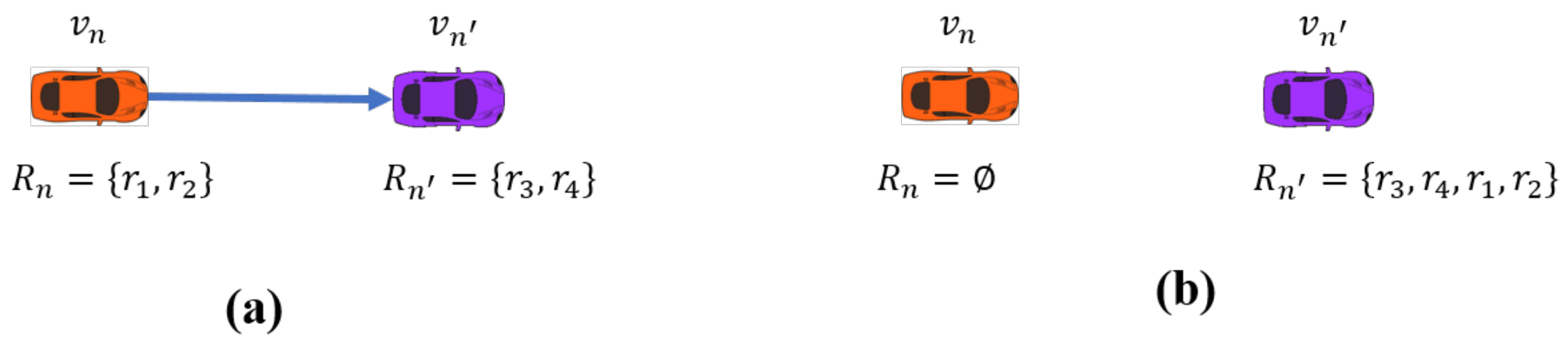

Figure 2: Matching of vehicles

Fig. 3 illustrates the GMOMatch algorithm along with a simple example. The algorithm consists of two steps (Fig. 3a). In the first step, we consider a bipartite graph to match requests with vehicles. Vehicles can be idle or enroute. The output of this step is one-to-one matching and creating a set of assigned vehicles (Definition 1). In the second step, we create a vehicle directed graph whose vertices are assigned vehicles. We then solve a maximum weight matching problem to match assigned vehicles with each other (Definition 2) and combine associated requests. The main algorithm as well as the step 2 are iterative and ends when some criteria are satisfied (see section 3.4). The proposed algorithm is graph-based because it takes advantage of a bipartite graph in Step 1 and a general graph in Step 2 to match the ride requests with vehicles.

We assume that the shared mobility service has a routing agent that can calculate the optimal path between two locations on a network using existing or predicted traffic conditions. This is a practical assumption, given the current level of sensorization in dense urban areas and the amount of probe vehicles on the network.

Figures 3b and 3c show a small example of how the GMOMatch algorithm works to assign requests to vehicles. In Fig. 3b, there are a set of requests and vehicles. The algorithm starts with solving a one-to-one matching problem in step 1 and assigns four requests (out of eight) to four vehicles. In the first iteration of step 2, by solving a maximum weight matching problem, vehicle 3 is matched with vehicle 2 which leads to combining requests 3 and 7 together in the vehicle 2. Similarly, in the second iteration of step 2, vehicle 1 is matched with vehicle 2 and its associate travel request 2 is combined with requests 7 and 3 in vehicle 2. Since four non-assigned requests are still left, the algorithm needs to move toward the second iteration (Fig. 3c). In step 1 of the second iteration, the same set of vehicles is considered and the remaining requests (four requests) assign to them. Finally, in step 2 of the second iteration, vehicle 3 is matched with vehicle 1, and requests 1 and 6 are combined in vehicle 1. Likewise, the algorithm continues until some criteria and constraints are met. In the following, the two steps of the GMOMatch algorithm will be explained in detail.

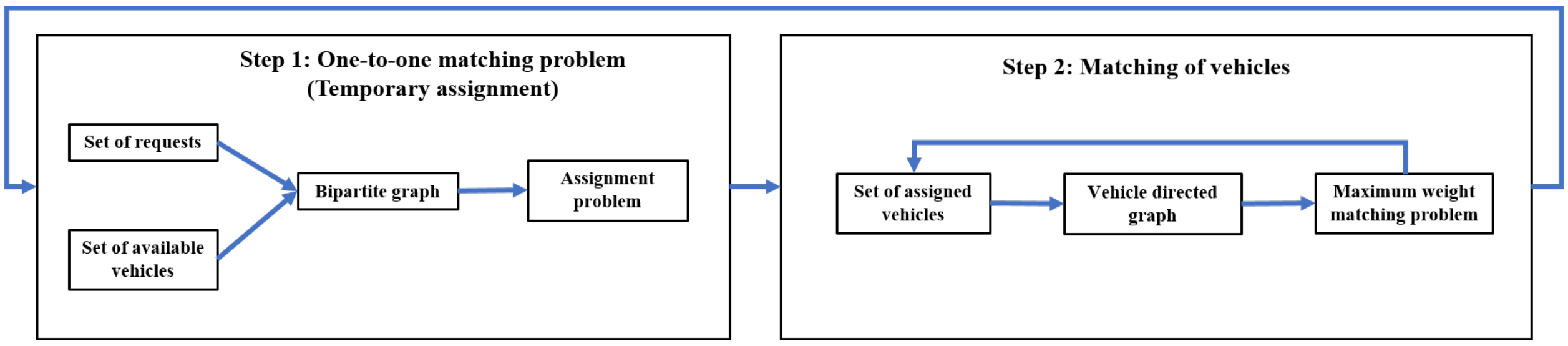

(a) GMOMatch algorithm

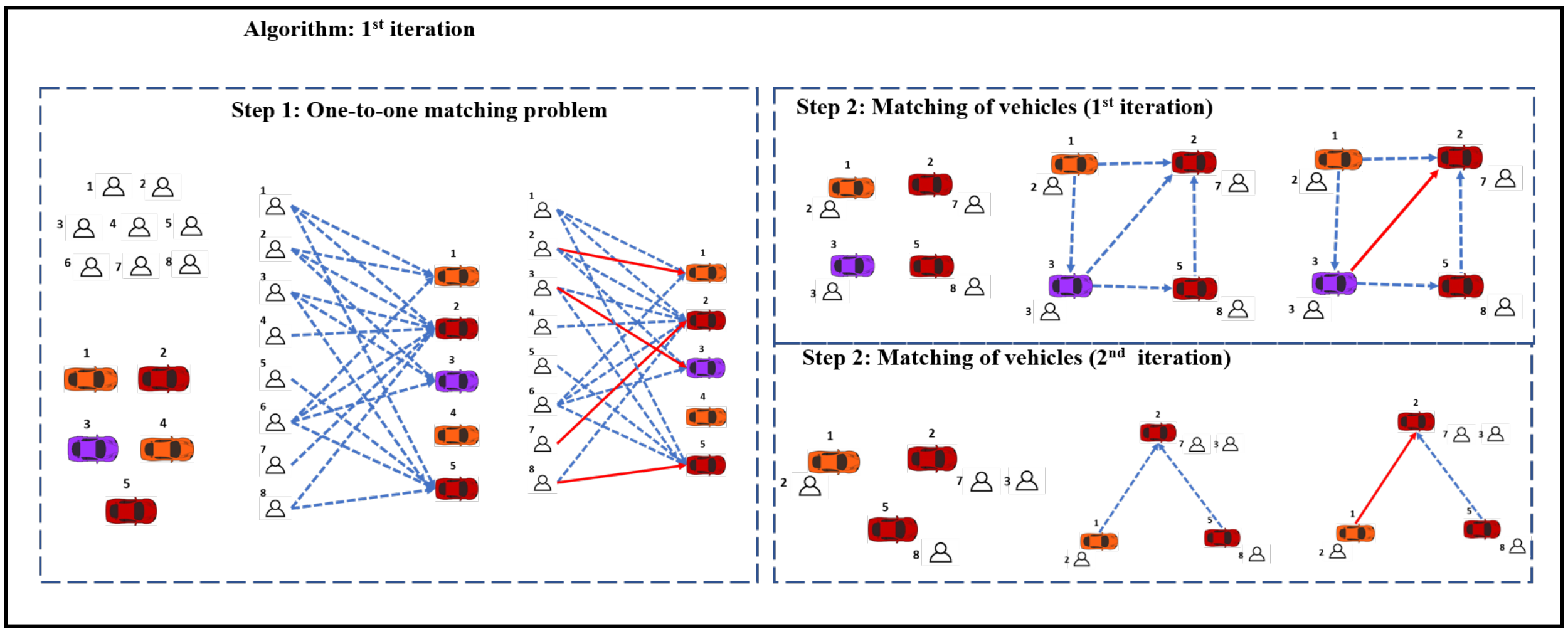

(b) First iteration of the algorithm

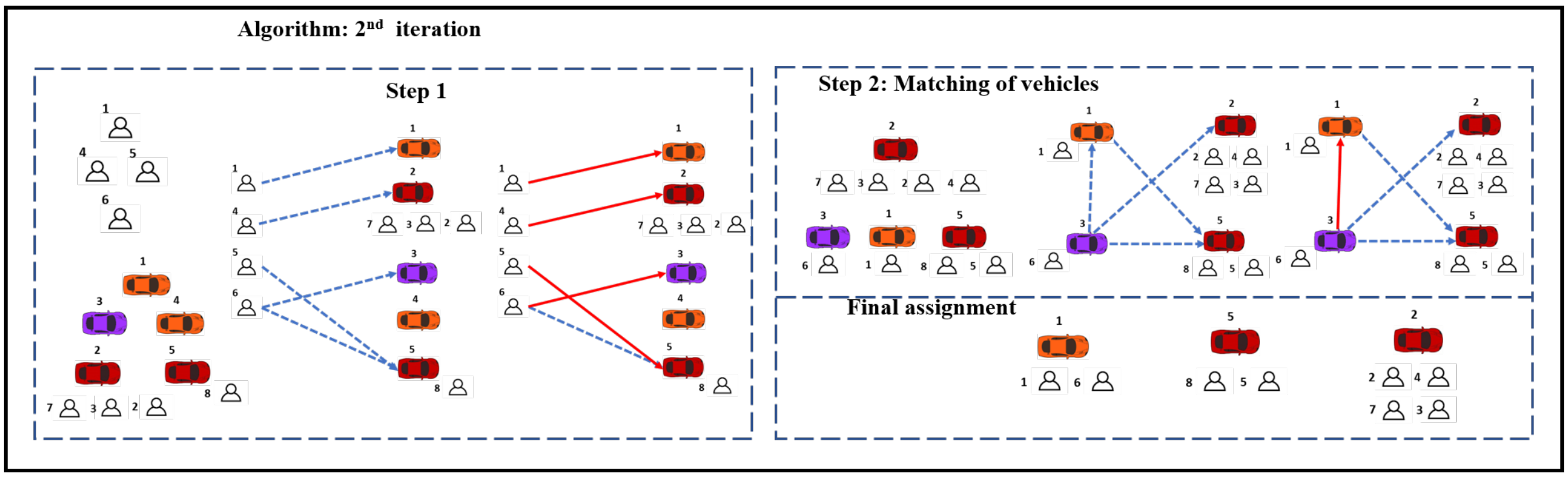

(c) Second iteration of the algorithm

Figure 3: Two-step iterative method

### 3.3. Step 1: One-to-one matching problem

The first step of the method is one-to-one matching problem, which can be represented by a bipartite graph $G = (I, L)$, where $I$ is the set of nodes, including all requests and vehicles $I = R \cup V$, and $L$ is the set of links. Link $l_{rv} \in L$ between request $r$ and vehicle $v$ exists if constraint set $Z$ is satisfied. Each link has a travel cost $c_{rv} \in C$ where $C$ is the set of travel costs and $c_{rv}$ is defined as the time duration needed by vehicle $v$ to serve both its already scheduled passengers and request $r$. The goal of the proposed one-to-one matching problem is minimizing vehicles' total travel time.

Let $M$ be the current location of vehicle $v \in V$, then $z_1$ and $z_2$ can be expressed mathematically as Eq. 1 and Eq. 2.

$$t + t_{rv}(M, O_r) \leq q_r \tag{1}$$

$$t + t_{rv}(M, O_r) + t_{rv}(O_r, D_r) \leq l_r \tag{2}$$

Where $t$ is the current time, $t_{rv}(M, O_r)$ is travel time from vehicle's current location to request's origin, and $t_{rv}(O_r, D_r)$ is travel time between request's origin and destination. As mentioned, Eq. 1 ensures that request $r$ would be picked up by vehicle $v$ not later than his latest departure time $q_r$, and Eq. 2 expresses that request $r$ would be dropped off not later than its latest arrival time $l_r$.

To create bipartite graph $G$, first, as a pre-processing, the capacity constraint $z_0$ is checked and all vehicles with at least one available seat are selected. we then specify the set of vehicles that can potentially serve each request ($V_r^f$ , $\forall r \in R$) and name it the set of feasible vehicles for request $r$. To this end, for each travel request $r$, a search space based on its flexibility $f_r$ is created. By assuming request $r$ as the first passenger to be picked up, all of the vehicles whose travel time from their current location to the request's origin are not greater than the request's flexibility are considered (Eq. 3). Vehicles can be idle or enroute. A vehicle is enroute if it has already been assigned to some passengers, heading to pick up/drop off locations. $V_r^f$ is empty if there are no feasible vehicles in the spatial vicinity of the request $r$. It is worth mentioning that creating a search space for the requests can significantly reduce the problem's computational complexity. Given $m$ number of requests and $n$ number of vehicles, to calculate travel time, the worst case is that for each request we consider all vehicles that is $O(mn)$ which scales linearly with the number of vehicles and requests. However, creating a search space reduces the number of feasible vehicles significantly. Let $k$ be the maximum number of feasible vehicles, and $k \ll n$. In this case, the computational complexity would be $O(mk)$.

$$t_{rv}(M, O_r) \leq f_r \tag{3}$$

**Proposition 1:** Vehicles violating Eq. 3 cannot serve request $r$.

**Proof:** According to $z_1$, we have the inequality $t + t_{rv}(M, O_r) \leq q_r$. On the other hand, from section 3.1, for each request $r$, $t \geq e_r$. Assuming that request $r$ is the first passenger to be picked up by vehicle $v$, $t_{rv}(M, O_r) > f_r$ that is the violation of Eq. 3. Thus, $t + t_{rv}(M, O_r) \geq e_r + f_r = q_r$ (from section 3.1) showing that $z_1$ is not satisfied and vehicle $v$ cannot serve request $r$.

After finding potential vehicles for each request, their travel costs and optimal travel paths are calculated. Each vehicle $v \in V$ has a current travel path (before considering request $r$) denoted by $\Lambda_v$ and updated travel path (with considering request $r$) denoted by $\Lambda_v^t$ that specifies which locations need to be visited to pick up/drop off. As mentioned, $c_{rv} \in C$ is the time duration of travel path of vehicle $v$, $\Lambda_v^t$, to serve its existing scheduled passengers as well as request $r$.

To obtain travel cost $c_{rv}$ and optimal travel path $\Lambda_v^*$, a vehicle routing problem, which in the literature is known as a single-vehicle DARP (Häme, 2011; Liu et al., 2015; Ho et al., 2018), needs to be solved. The objective is to minimize vehicle $v$'s travel time subject to the constraints $z_1$ and $z_2$. To do so, we propose a function whose inputs are $\Omega_r$, $\Lambda_v$, and $\Theta_{P_v}$ where $\Omega_r$ is spatiotemporal information of request $r$ (e.g. origin, destination, time window), and $\Theta_{P_v}$ represnets the spatiotemporal information of already scheduled passengers $P_v$. This function is proposed as Eq. 4.

$$(c_{rv}, \Lambda^t_v) = PathCost(\Omega_r, \Lambda_v, \Theta_{P_v}) \quad (4)$$

To solve the proposed single-vehicle DARP (adding a new request $r$ to current vehicle path $\Lambda_v$), all possible combinations are enumerated for vehicles that already have at most two scheduled passengers (four locations in their path to pick up/drop off). For vehicles with more than two scheduled passengers, as in Alonso-Mora et al. (2017) and Simonetto et al. (2019), an insertion heuristic method (Algorithm 1) is used based on which new request's pick up and drop off locations are inserted, while the current order of the schedule of $\Lambda_v$ is kept. For instance, the current travel path of vehicle $v$ is $tour = (+p_2, -p_1, -p_2)$ which means passenger 2 pick up , passenger 1 drop off, and passenger 2 drop off. Request 3 can be added as $newtour = (+r_3, -r_3, +p_2, -p_1, -p_2)$, $newtour = (+r_3, +p_2, -p_1, -p_2, -r_3)$, $newtour = (+p_2, -p_1, +r_3, -p_2, -r_3)$, and *etc*. For each *newtour*, constraints $z_1$ and $z_2$ for both new request $r$ and on-board passengers are checked. In the case where they satisfied, *newtour* is considered a feasible travel path. In algorithm 1, $k$ enumerates feasible vehicle travel paths and *maxk* represents the total number of feasible travel paths. The time duration of each feasible *newtour* is then calculated and among feasible travel paths, the one with minimum travel time is chosen. Given $s$ be the number of scheduled locations (origin and destination) in the vehicle tour, based on the proposed insertion method there would be $(s + 1)$ spots for the new request's origin and destination. Thus, the computational complexity of the insertion method for one request is $O(s^2)$ and for the entire bipartite graph with $mk$ edges is $O(mks^2)$.

The one-to-one matching problem presented here can be mathematically formulated as an integer programming model (5). The decision variable $x_{rv}$ is 1 if vehicle $v$ and request $r$ match with each other and 0 otherwise. The objective function (Eq. 5a) aims at minimizing the total travel time of the vehicles. Constrains 5b and 5c ensure that each vehicle/request is matched with one request/vehicle (if symmetric $|R| = |V|$). Due to the structure of constraints in linear assignment problems which is totally unimodular, Lemma 1 (De Giovanni, 2020), the binary constraint $x_{rv} \in \{0, 1\}$ can be relaxed and expressed as Constraint 5d.

**Algorithm 1**

$tour \leftarrow \Lambda_v$
$L \leftarrow length(tour)$
$1 \leftarrow k$
tour($w : y$) returns the $w$-th index to $y$-th index
**for** i=0:L **do**
  $newtour \leftarrow [\, tour(1 : i)\;\; O_r\; tour(i + 1 : end)]$
  $M \leftarrow length(newtour)$
  **for** j=i+1:M **do**
    $newtour \leftarrow [\, newtour(1 : j)\; D_r\;\; newtour(j + 1 : end)]$
    **if** $newtour$ satisfies $z_1$ and $z_2$ **then**
      $c^k_{rv}$ ← time duration of $newtour$
      $newtour^k \leftarrow newtour$
      $k + 1 \leftarrow k$
    **end if**
  **end for**
**end for**
$k* \leftarrow arg\, min\{c^k_{rv}\}_{\{k=1,...,maxk\}}$
**return** $c_{rv} = c^{k*}_{rv} \quad \Lambda^t_v = newtour^{k*}$

$$\min \sum_{r\in R} \sum_{v\in V} c_{rv}x_{rv} \tag{5a}$$

$$\sum_{r\in R} x_{rv} = 1 \qquad \forall v \in V \tag{5b}$$

$$\sum_{v\in V} x_{rv} = 1 \qquad \forall r \in R \tag{5c}$$

$$0 \leq x_{rv} \leq 1 \tag{5d}$$

**Lemma 1.** *Let A be a matrix with all entries in* $\{0, +1\}$ *such that every column of A has at most two entries of value 1. If the rows of A can be partitioned into two sets* $V1$ *and* $V2$ *such that every column of A has at most one entry of value 1 in the rows of* $V1$ *and at most one entry of value 1 in the rows of* $V2$*, then A is completely unimodular.*

*Proof.* Let $B$ be a $k \times k$ square submatrix of $A$. It is shown by induction on $k$ that $det(B) \in \{0, +1, 1\}$. If $k = 1$, then $B$ has a single entry that, by assumption, is 0 or 1, and therefore $det(B) \in 0, 1$. Now take $k \geq 2$ and assume by induction that every $(k - 1) \times (k - 1)$ square submatrix of $A$ has determinant 0, +1 or 1. Note that every column of $B$ has at most two entries of value 1. Three cases can be considered.

a) $B$ has at least one all-zero column. In this case $det(B) = 0$.

b) $B$ has at least one column with exactly one entry equal to 1. Let us assume that the $j-th$ column of $B$ has a single entry of value 1, say the entry in row $i$. By Laplace rule, if $B$ is the submatrix of $B$ obtained by removing the $i-th$ row and the $j-th$ column, then $det(B) = (-1)^{(i+j)}det(B^{t})$. By induction, $det(B) \in \{0, +1, 1\}$, and therefore $det(B) = (-1)^{(i+j)}det(B^{t}) \in \{0, +1, 1\}$.

c) Each column of $B$ has precisely two entries of value 1. In this case, by assumption every column of $A$ has one entry of value 1 in the rows of $V1$ and one entry of value 1 in the rows of $V2$. Then the sum of the rows of $B$ in $V1$ minus the sum of the rows of $B$ in $V2$ is the zero vector. This implies that the rows of $B$ are linearly dependent, and therefore $det(B) = 0$. □

Fig. 4 shows the incidence matrix of constraints in assignment problem 5. The rows of matrix are partitioned into two parts of $r \in R$ and $v \in V$ that represent ride requests and vehicles, respectively. Each column can have at most two 1 values if link $l_{rv}$ between $r$ and $v$ exists. Thus, according to Lemma 1, this matrix and as a result, the assignment problem 5 is totally unimodular.

| | | $r_1v_1$ | $r_1v_2$ | $\cdots$ | $r_1v_n$ | $r_2v_1$ | $\cdots$ | $r_mv_n$ |
|---|---|---|---|---|---|---|---|---|
| $r \in R$ | $r_1$ | 1 | 0 | $\cdots$ | 1 | 0 | $\cdots$ | 0 |
| | $r_2$ | 0 | 0 | $\cdots$ | 0 | 1 | $\cdots$ | 0 |
| | $\vdots$ | $\vdots$ | $\vdots$ | $\vdots$ | $\vdots$ | $\vdots$ | $\vdots$ | $\vdots$ |
| | $r_m$ | 0 | 0 | $\cdots$ | 0 | 0 | $\cdots$ | 0 |
| $v \in V$ | $v_1$ | 1 | 0 | $\cdots$ | 0 | 1 | $\cdots$ | 0 |
| | $v_2$ | 0 | 0 | $\vdots$ | 0 | 0 | $\vdots$ | 0 |
| | $\vdots$ | $\vdots$ | $\vdots$ | $\vdots$ | $\vdots$ | $\vdots$ | $\vdots$ | $\vdots$ |
| | $v_n$ | 0 | 0 | $\cdots$ | 1 | 0 | $\cdots$ | 0 |

Figure 4: Incidence matrix of assignment problem 5

To solve the assignment problem 5, we use Hungarian algorithm, which is considered one of the common and effective approaches for such problems.

The output of the one-to-one matching problem here is the matching of requests with vehicles and pick up/drop off scheduling(travel path).

*3.4. Step 2: Matching of vehicles*

The second step of the algorithm is iterative. In the first iteration, the input is the output of step 1 while for the following iterations, the input is the output of the previous iteration. Let $V^t = \{v_1, v_2, ..., v_n\} \subseteq V$ be the set of assigned vehicles from step 1. $P_v^t$ represents the updated $P_v$ for assigned vehicles and is $P_v^t = R_v \cup P_v$.

We define a directed graph $G_v = (I^t, L^t)$ where $I^t$ is the set of nodes representing assigned vehicles $V^t$ and $L^t$ is the set of directed links. Hereafter in this study, graph $G_v$ is called vehicle graph. To create vehicle graph $G_v$, first we need to determine which nodes can be connected to each other. Fig. 5 showcases an example of the potential nodes that each node can be connected with. Fig. 5a represents a bipartite graph with four requests and seven vehicles. From step 1, From Step 1, we know that $V_r^f$ can be empty. if this is the case, the request is moved to next time window for matching. In case of $V_r^f$ not to be empty, it is processed in this time window. In Fig. 5b, through one-to-one assignment in step 1, requests and vehicles are matched together. As mentioned, assigned vehicles $V^t = \{v_1, v_4, v_5, v_6\}$ constitute the nodes of vehicle graph (Fig. 5c,). Each vehicle $v \in V$ in vehicle graph can only be connected to the set of feasible vehicles of request $r$. For instance, in Fig. 5*c*, $v_1$ can only be connected to $v_4$ and $v_6$ because the set of feasible vehicles of request $r_2$ in request-vehicle bipartite graph (Fig. 5a) is $V_2^f = \{v_1, v_3, v_4, v_6\}$. Likewise, $v_6$ can be connected to $v_1$ and $v_5$ because $r_3$' feasible vehicle set is $V_3^f = \{v_1, v_3, v_5, v_6\}$. However, $v_5$ in this vehicle graph as seen cannot be connected to any other vehicle since $r_1$' feasible vehicle set is $V_1^f = \{v_2, v_3, v_5, v_7\}$.

Given the number of assigned vehicles $n^t$, in the worst case, when each vehicle can be connected to all other vehicles, the complexity is $O(n^{t2})$. However, because from step 1 (3.3) each request is connected to $k$ feasible vehicles, in the worst case, each vehicle can be connected to $(k - 1)$ vehicles. Thus, the computational complexity is $O(n^t k)$.

After determining feasible nodes for each node in the vehicle graph, we need to create the set of directed links $L^t$. A directed link $l_{vv'}^t \in L^t$ between any two assigned vehicles $v$ and $v^t$ $(v, v^t \in V^t)$ exists if constraints $Z^t$, including $z_1^t$, $z_2^t$, and $z_3^t$ are satisfied. These constraints are defined as follows: $z_1^t$) only idle vehicles can be matched with other vehicles (idle/enroute) $z_2^t$) assigned vehicles with less occupants can be matched with assigned vehicles with more occupants $z_3^t$) the size of assigned requests set $|R_v|$ associated with assigned vehicle $v \in V^t$ should be less than or equal to current available capacity of other assigned vehicles in the vehicle graph. The logic behind $z_1^t$ is that enroute vehicles have already some on-board passengers, and rerouting passengers to another vehicle in the hope of travel time reduction may be considered impractical (e.g. transferring from one vehicle into another is not acceptable by the passengers). The reasoning behind $z_2^t$ is delivering a higher-quality solution for vehicle travel path which will be explained further. Finally, constraint $z_3^t$ ensures that the vehicle has enough available capacity.

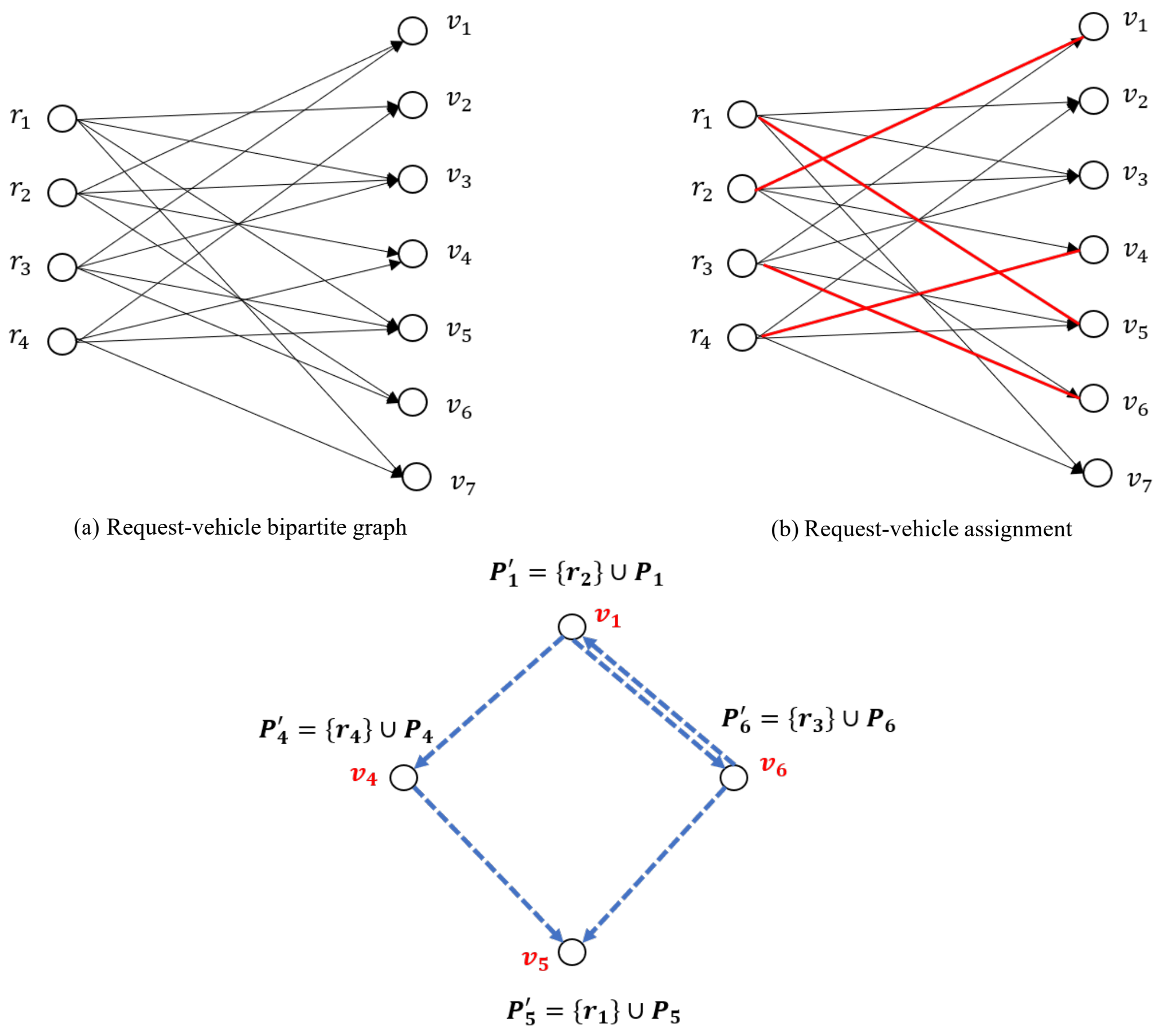

(a) Request-vehicle bipartite graph

(b) Request-vehicle assignment

(c) Prospective links in vehicle graph

Figure 5: Vehicle graph potential links

Fig. 6 shows an example to clarify these constraints. Consider vehicle graph in Fig. 6a in which $\{v_1, v_2, v_4\}$ are idle and $\{v_3, v_5\}$ are enroute. It is assumed that the total capacity of vehicles is six, and two vehicles of $v_3$ and $v_5$ each has four occupants($|P_3| = 4$ and $|P_5| = 4$).

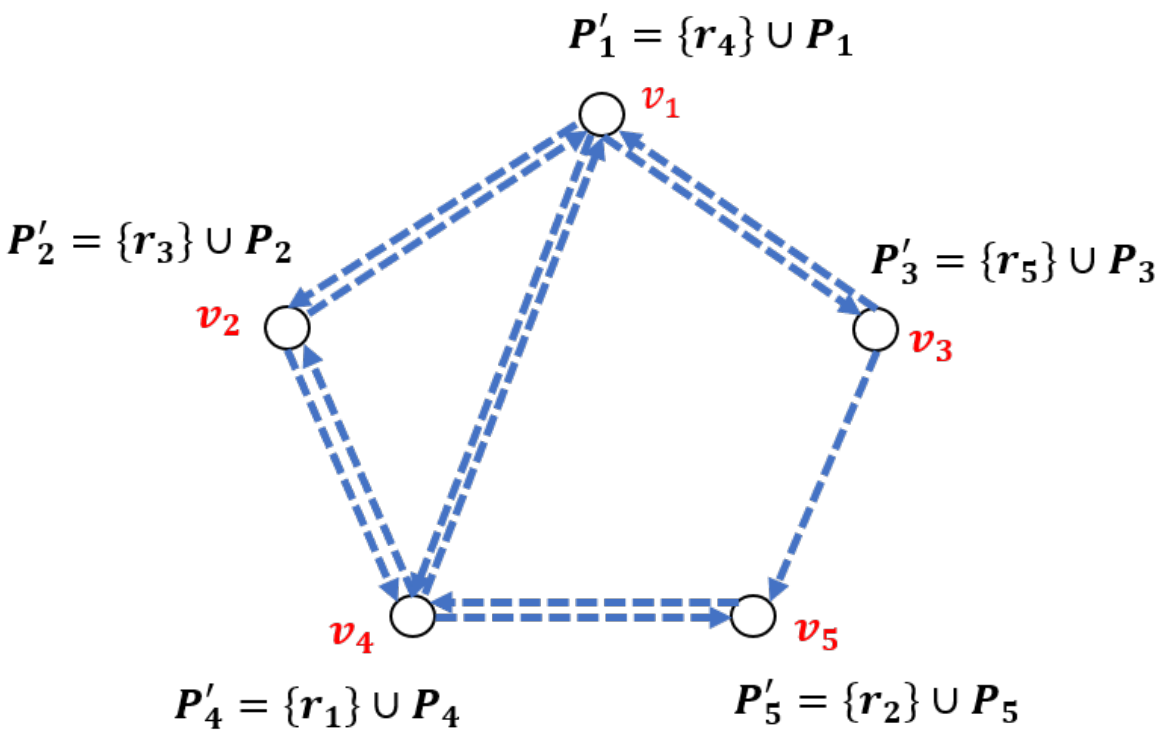

*(a)* First iteration before checking $z_1$

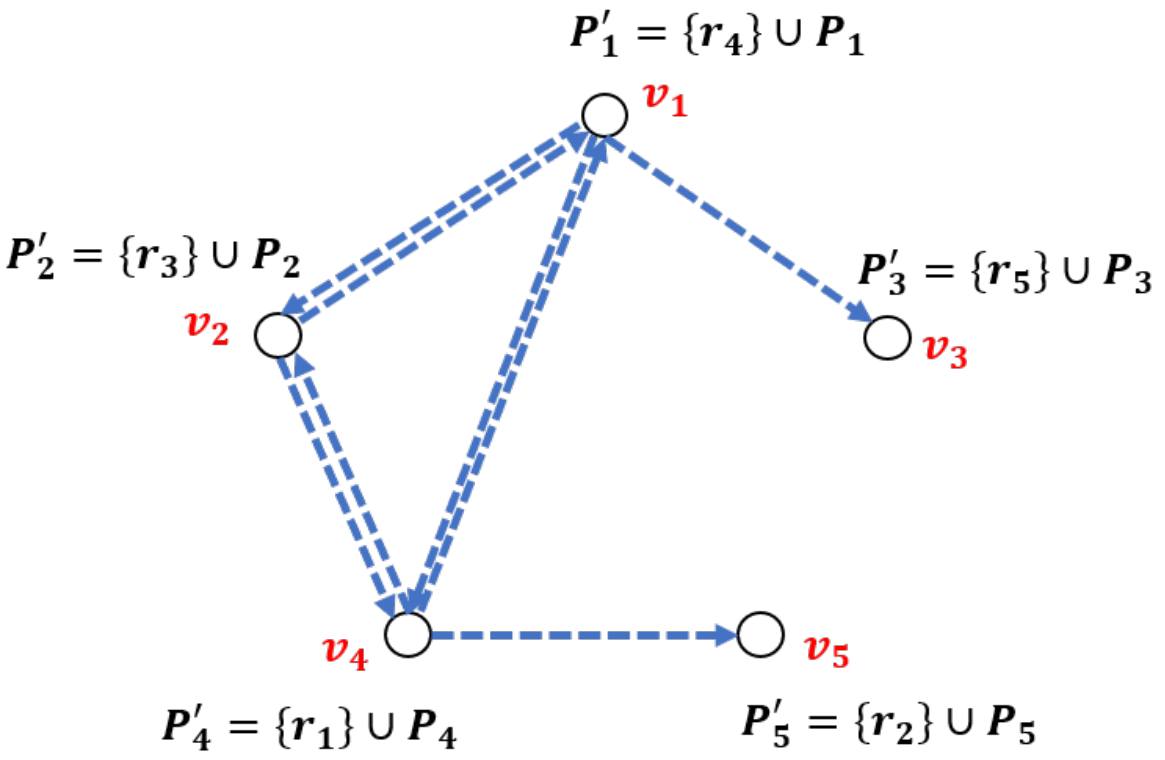

*(b)* First iteration after checking $z_1$

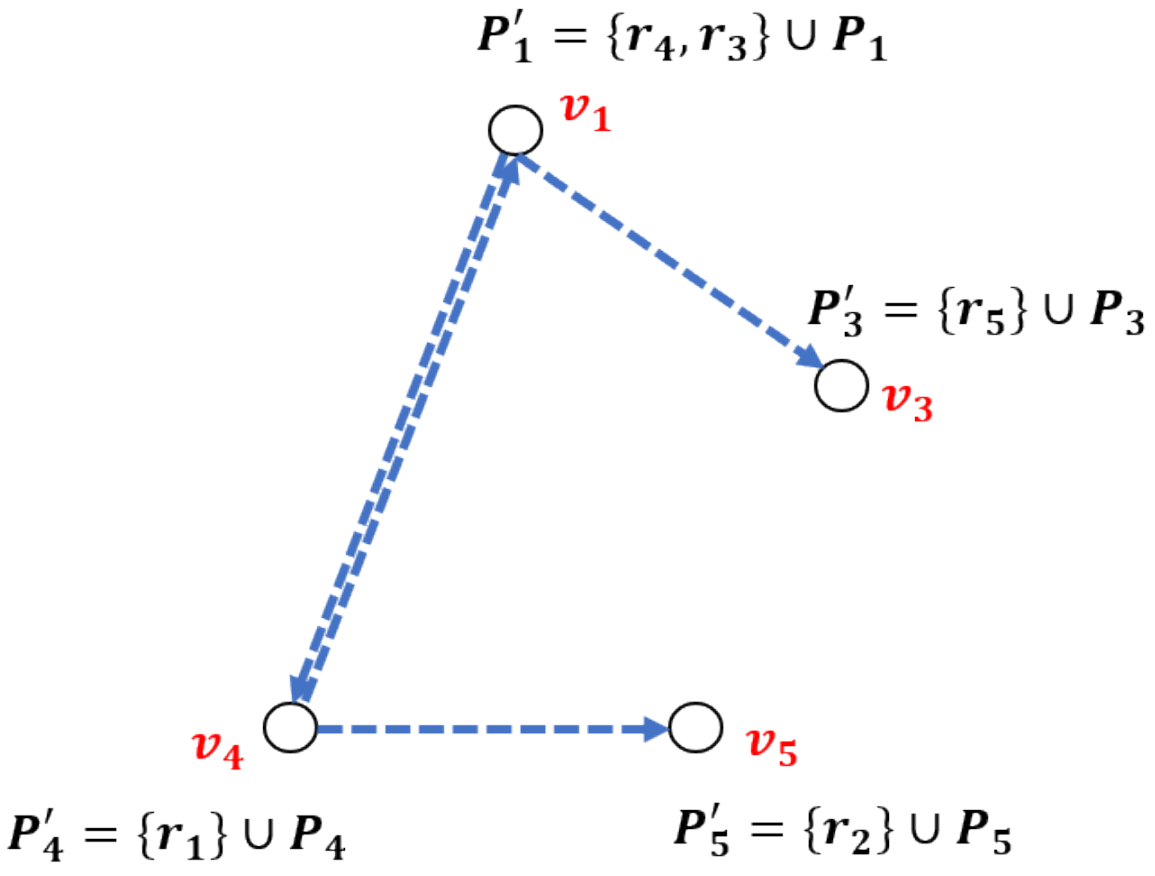

*(c)* Second iteration before checking $z_2$

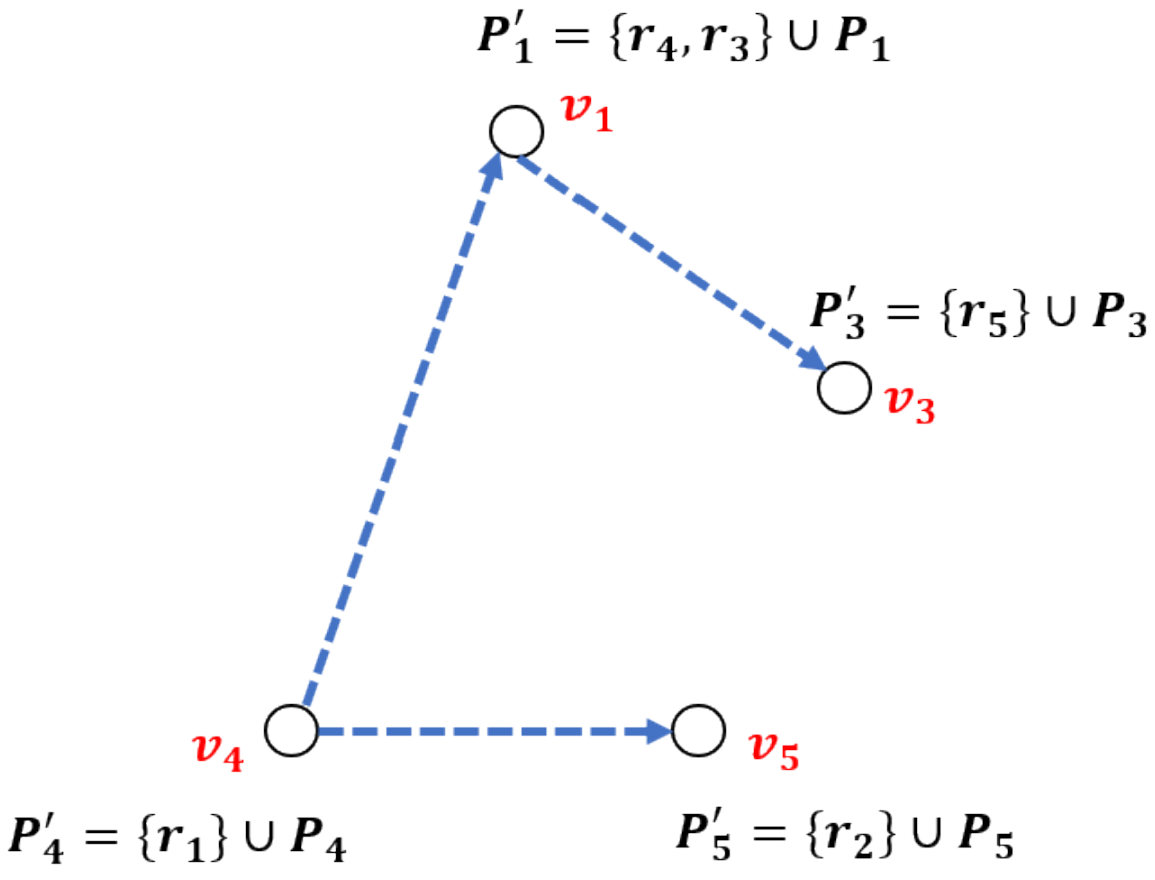

(d) Second iteration after checking $z_2$

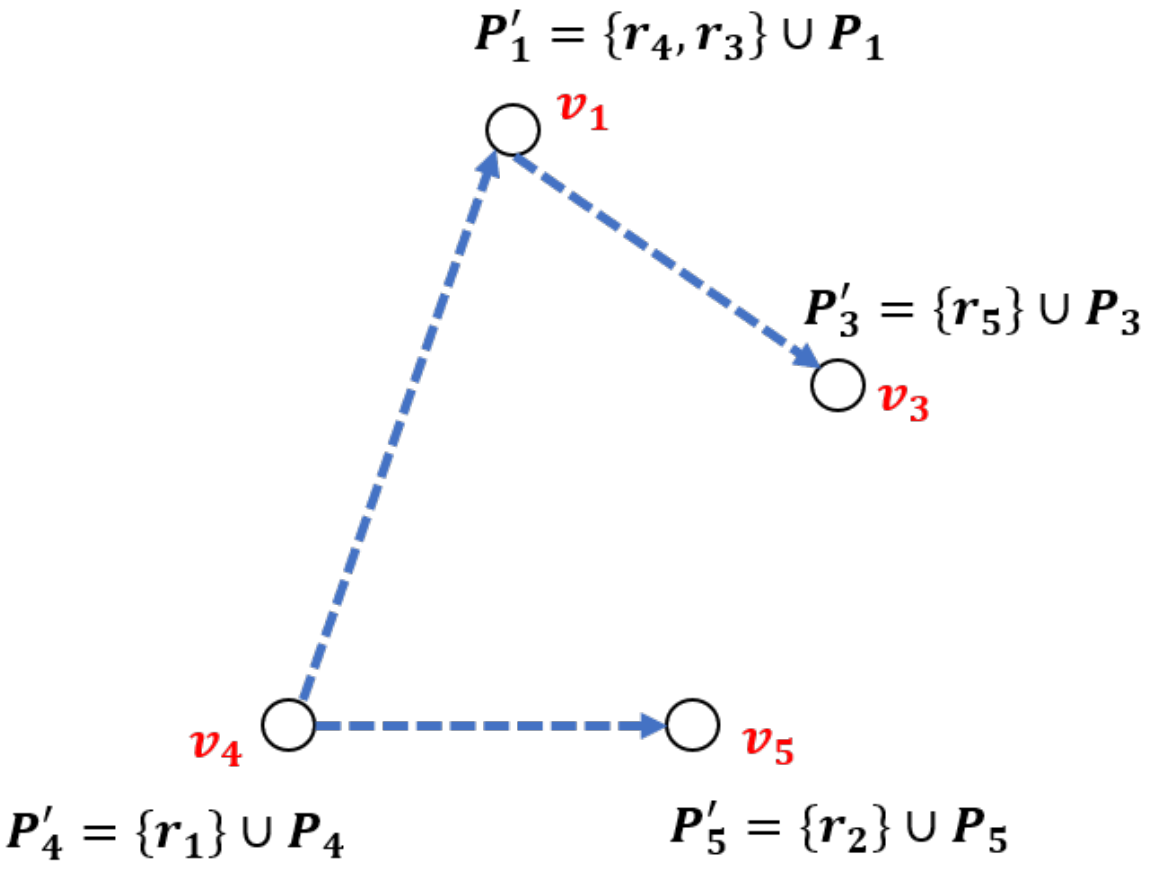

*(e)* Second iteration before checking $z_3$

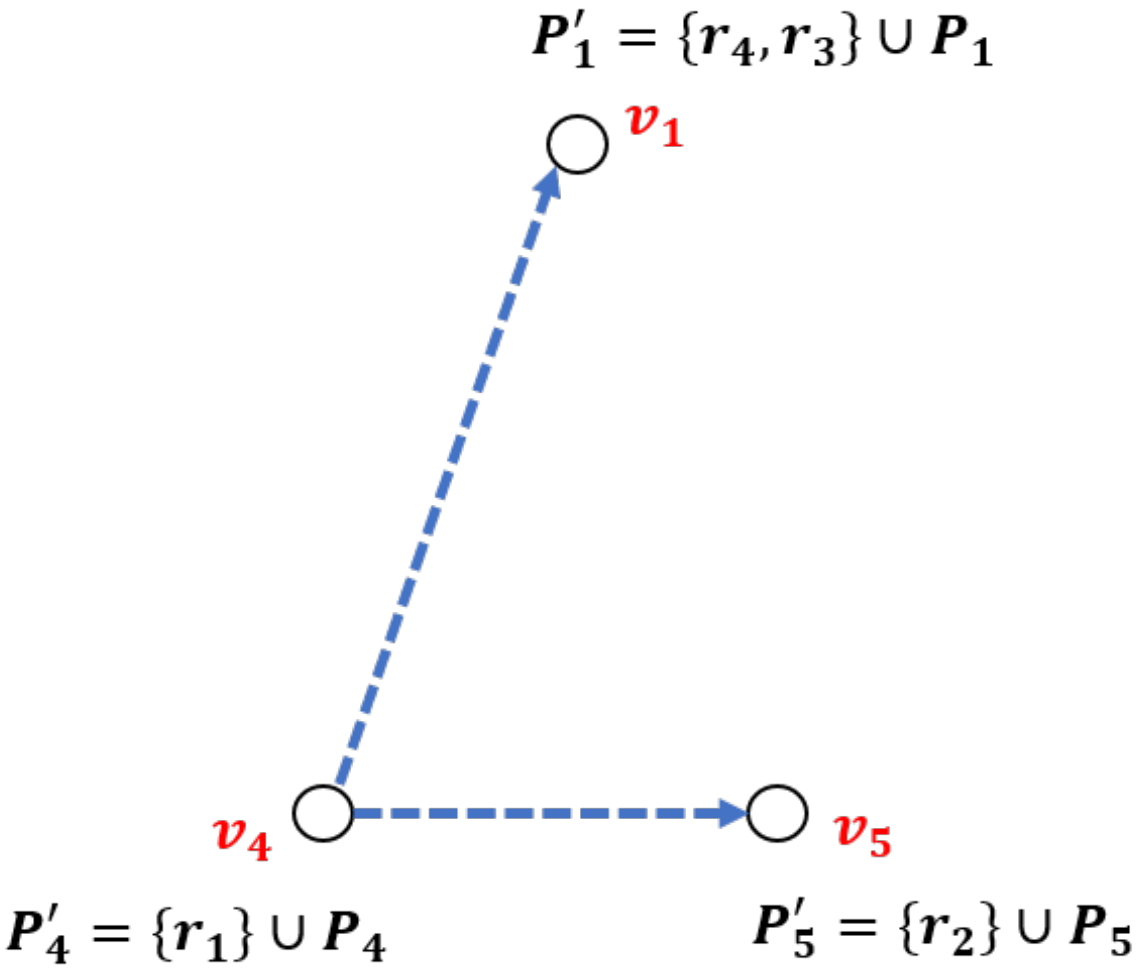

*(f)* Second iteration after checking $z_3$

Figure 6: Vehicle graph: creating set of links

According to constraint $z_1^{\dagger}$, idles vehicles $\{v_1, v_2, v_4\}$ can be matched with any other vehicles while $\{v_3, v_5\}$ as enroute vehicles cannot be matched. Fig. 6b represents the modified vehicle graph based on $z^t_1$. Assuming that in the first iteration of step 2, $v_2$ is matched with $v_1$, the vehicle graph in the second iteration is formed as Fig. 6c . In this Figure, based on the constraint $z^t_2$, $v_4$ can be matched with $v_1$ ($|P^t_4| = 1 \leq |P^t_1| = 2$), $v_1$ can be matched with $v_3$ ($|P^t_1| = 2 \leq |P^t_3| = 5$) and $v_4$ can be matched with $v_5$ ($|P^t_4| = 1 \leq |P^t_5| = 5$). The altered vehicle graph can be seen in Fig. 6d. Constraint $z^t_3$ expresses that vehicles should have enough empty seat. For instance, in the obtained vehicle graph shown in Fig. 6e, the set of assigned requests of $v_1$ ($R_1 = \{r_4, r_3\}$) which has two request members cannot be matched with $v_5$ because it only has one empty space ($|R_1| = 2 \geq cap^t_5 = 1$) while $v_4$ can be matched to other vehicles of $v_1$ and $v_5$.

In addition to these three constraints, constraints $z_1$ and $z_2$ need to be satisfied. Notice that constraints should be checked for all members of assigned requests set $R_v$. As in Step 1, a routing function is used through which these two constraints are checked and optimal path and associated cost is obtained. This routing function is as Eq. 6.

$$(c^t_{vv}, \Lambda^*_v) = PathCost(\Omega_{R_v}, \Lambda^t_v, \Theta_{P_v}) \tag{6}$$

Where $\Omega_{R_v}$ is the spatiotemporal information associated with the members of assigned requests set $R_v$, $\Lambda^t_v$ is the updated travel path of vehicle $v$, $\Theta_{P_v}$ represents the spatiotemporal information of updated scheduled passengers $P^t_v$, $c^t_{vv}$ is the travel cost associated with the directed link $l^t_{vv}$ and $\Lambda^*_v$ is the optimal travel path of vehicle $v^t$. To calculate travel cost and optimal travel path, we modified the insertion heuristic method presented in Step 1 (e.g. Algorithm 1) to propose algorithm 2.

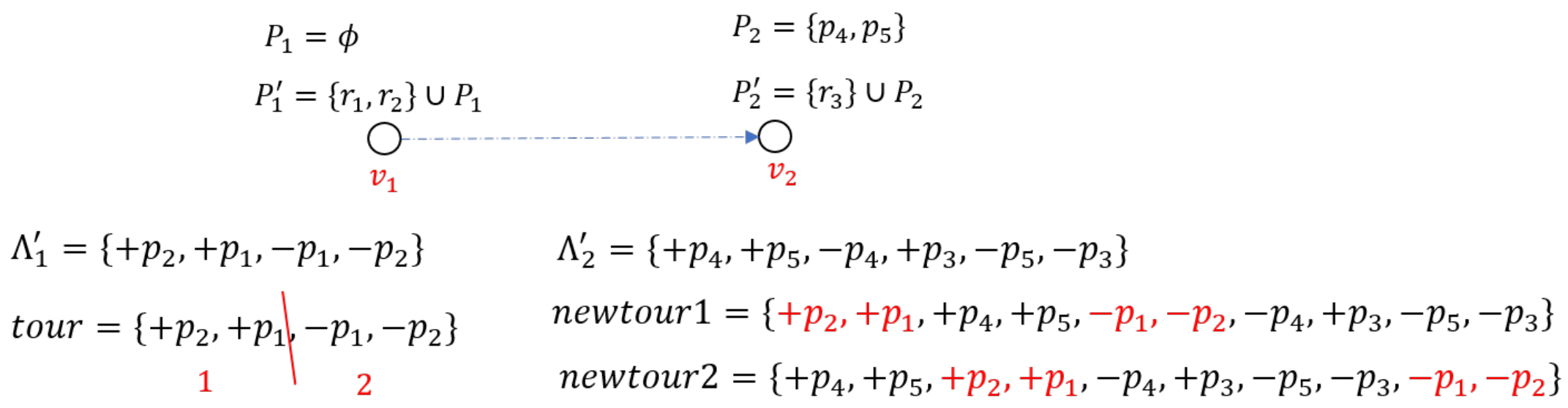

Figure 7: Creating new travel path

To illustrate this algorithm, consider two vehicles in Fig. 7 which $v_1$ is idle and $v_2$ is enroute.

To match $v_1$ and $v_2$ with each other and assign $R_1 = \{r_1, r_2\}$ to $v_2$, the current travel path of $v_1$, which is represented by $tour = \Lambda_1^t$, is divided into two parts from the middle. To create a new travel path for $v_2$, each part is added to $\Lambda_2^t$, while the current sequence of pick up/drop off is remained. Notice that to create a new path, part 1 of *tour* should be placed before part 2. *newtour*1 and *mewtour*2 in Fig. 7 show two examples of new travel path of $v_2$. For each new travel path, first the constraints $z_1$ and $z_2$ are checked and then the time duration of each *newtour* is calculated. Finally, among different travel path combinations, the one with minimum travel time is chosen. Based on the set of travel costs $c_{vv}^t \in C^t$ acquired from Eq. 6, the set of directed links $l_{vv}^t \in L^t$ of vehicle graph is determined.

Similar to the previous insertion method, given $s$ be the number of scheduled locations in the vehicle tour, there will be $(s + 1)$ spots for the new riders' Part 1 and 2. Thus, the insertion method's computational complexity for one request is $O(s^2)$ and for the entire vehicle graph with $n^t k$ edges is $O(n^t k s^2)$.

**Algorithm 2**

$tour_1 \leftarrow \Lambda_v^t$
tour breaks into two parts
$1stPart \leftarrow$ tour first part
$2ndPart \leftarrow$ tour second part
$tour_2 \leftarrow \Lambda_v^t$
$L \leftarrow length(tour_2)$
$L^t \leftarrow length(1stPart)$
$1 \leftarrow k$
tour$(w : y)$ returns the $w$-th index to $y$-th index
**for** i=0:L **do**
  $newtour \leftarrow [\,tour_2(1 : i)\ 1stPart\ \ tour_2(i + 1 : end)]$
  $M \leftarrow length(newtour)$
  **for** j=i+$L^t$:M **do**
    $newtour \leftarrow [\,newtour(1 : j)\ 2ndPart\ \ newtour(j + 1 : end)]$
    **if** *newtour* satisfies the constrains $Z$ **then**
      $c_{vv}^{t(k)} \leftarrow$ time duration of *newtour*
      $newtour^k \leftarrow newtour$
      $k + 1 \leftarrow k$
    **end if**
  **end for**
**end for**
$k* \leftarrow arg\,min\{c_{vv}^{t(k)}\}_{\{k=1,\ldots,maxk\}}$
**return** $c_{vv}^t = c_{vv}^{(k*)}$ $\Lambda_v^* = newtour^{k*}$

Fig. 8 illustrates constraint $z_2^t$ with an example. There are two idle vehicles of $v_1$ and $v_2$

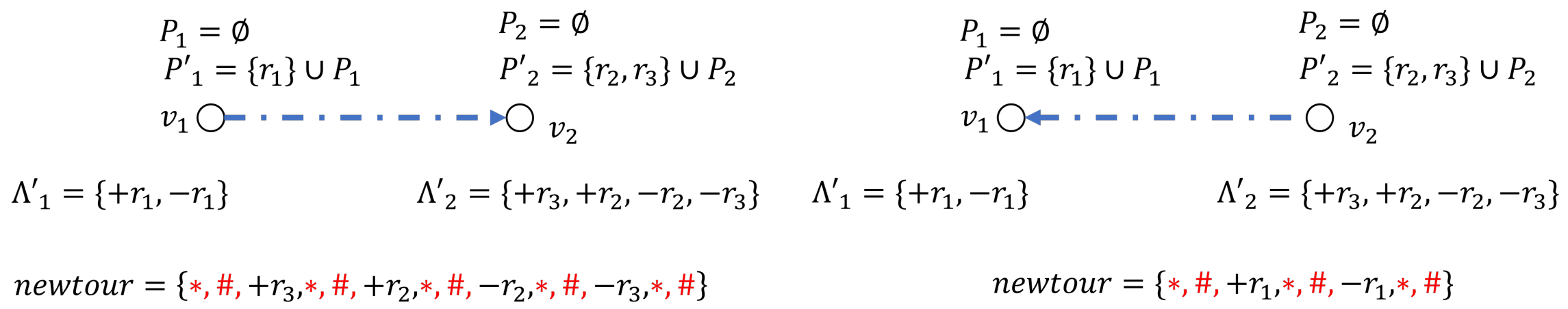

(a) $v_1$ is matched with $v_2$ (b) $v_2$ is matched with $v_1$

Figure 8: Constraint $z_2^I$

such that $R_1 = \{r_1\}$ and $R_2 = \{r_2, r_3\}$. To match $v_1$ with $v_2$ (Fig. 8a) and create $v_2$'s travel path, according to algorithm 2, there are five spots for placing $+r_1$ in *newtour* (red stars) and five spots for $-r_1$ (red sharps) which in total 25 combinations need to be addressed, denoted by set $A$, to find the best solution. However, to match $v_2$ with $v_1$ (Fig. 8b) and create $v_1$'s travel path, there are three spots for placing $\{+r_3, +r_2\}$ in *newtour* (red stars) and $\{-r_2, -r_3\}$ (red sharps) which creates 9 different combinations, denoted by set $A^t$. Given that more number of combinations are created and addressed in the former, the obtained solution from the former has most likely lower tour time.

In the obtained directed vehicle graph, the purpose is to find a matching to minimize the total travel cost. This problem is equivalent to the maximum weight matching problem in graph theory where the purpose is to find a matching in a weighted graph, where the sum of weights is maximized. To solve our problem, first it is converted into the standard form of maximum weight matching problem by multiplying the travel costs by minus one. Moreover, without the loss of generality, we assume that the vehicle graph is undirected. We can impose this assumption because total weights are independent from the link directions. To solve the maximum weight matching problem, we use Edmonds' algorithm (Saunders, 2013). The computational complexity of this algorithm is $O(|V^t|^3)$, where $V^t$ is the set of assigned vehicles in the vehicle graph.

As Step 2 of the algorithm is iterative, it continues until the set of directed links in vehicle graph is empty ($L^t = \varnothing$). This occurs when at least one of the constraints of $Z^t$ is not satisfied. The algorithm stops when one of these conditions is satisfied: (1) set of travel requests is empty ($R = \varnothing$), or (2) set of links in bipartite graph $G$ (step 1) is empty ($L = \varnothing$), that occurs when one of the constraints of $Z$ is not satisfied (e.g. no available vehicles exist ).

*3.5. Computational comparison*

In the Simonetto algorithm Simonetto et al. (2019): (1) context mapping determines the potential vehicles for each request, (2) to specify a vehicle's travel path, a DARP with an insertion heuristic method is used, (3) an ILP to assign requests to vehicles is solved, and (4) a rebalancing strategy for the vehicles is used. In the GMOMatch algorithm, Step 1 includes all stages of the Simonetto algorithm except for rebalancing. We create a search space to determine potential vehicles for each request, which is equivalent to the context mapping. To specify a vehicle's travel path, we use a heuristic insertion method with the computational complexity of $O(s^2)$ for one request and $O(mks^2)$ for the entire bipartite graph which is a similar complexity as that of Simonetto. To solve the ILP with $N \times N$ cost matrix, GMOmatch uses Hungarian and Simonetto uses Auction algorithm—both having the complexity of $O(N^3)$. Unlike the Simonetto, which does not combine requests, the GMOMatch combines requests in its second step. In Step 2 of the GMOMatch, the computational complexity for the entire vehicle graph is $O(n^t k^t s^{t2})$. Moreover, the complexity of maximum weight matching problem in a general graph is $O(V^2 E)$ where $V$ represents number of vertices and $E$ is number of edges. Thus, the computational complexity of the entire algorithm is $O(max\{mks^2, N^3, n^t k^t s^{t2}, V^2 E\})$, which in the worst case is similar to the Simonetto's algorithm.

In the following, two most common cases in congested networks are addressed. Given $m$ number of requests, $n$ number of vehicles, $k$ maximum number of vehicles each request can be connected with, $s$ number of spots in vehicle travel tour, and $k, s \ll m, n$, if $m \geq n$, in step 1 of the algorithm a $N \times N$ assignment problem, $N = m$, should be solved. In the worst case, each vehicle has been assigned a request. In such case, the number of assigned vehicles in step 2 is $n = n^t$ which is also equal to $V = n^t$ in vehicle graph. Given each vehicle can be connected with at most $k^t$ vehicles, $s^t$ number of spots in vehicle travel tour in step 2, and $k^t \ll n^t$, $E = kn^t$. Thus, the $max\{mks^2, m^3, nk^t s^{t2}, n^3\}$ is $m^3$ which is equal to $max\{mks^2, m^3\} = m^3$ in the Simonetto's algorithm. Moreover, if $m < n$, in step 1 for the assignment problem $N = n$. In the worst case, all of the vehicles have been assigned a request. Thus, the number of assigned vehicles in step 2 is equal to $n^t = m = V$ and $E = k^t m$. Therefore, the $max\{mks^2, n^3, mk^t s^{t2}, m^3\}$ is $n^3$ which is equal to $max\{mks^2, n^3\} = n^3$ in the Simonetto's algorithm.

## 4. Case Study and Results

In this section, we briefly introduce the study area and explain how we synthesized the demand. The parameter settings for the simulation of the GMOMatch on micro-traffic simulator are described. To evaluate the performance of the GMOMatch, we compare it with the Simonetto's algorithm and discuss on the obtained results. Finally, to assess how changing key variables and GMOMatch parameters affect the performance of the algorithm, a detailed sensitivity analysis is conduced.

### *4.1. Case Study Implementation*

We considered road network of Downtown, Toronto as the study area. This network that faces recurrent congestion during morning and afternoon peak periods was chosen because we had access to the data of this network. Fig.9 presents the network, bounded by 3.14km x 3.31km, which consists of 268 nodes/intersections and 839 links.

We implemented the GMOMatch algorithm and the road network of Downtown, Toronto in MATLAB and applied them on an in-house agent-based micro-traffic simulator (Djavadian and Farooq, 2018). The dynamic demand loading period in this study was 7:45am-8:00am (15 minutes) in the morning peak period. The demand used in this study is time-dependent exogenous Origin-Destination (OD) demand matrices for the year 2018, which is based on 5 minutes intervals and were obtained from the Transportation Tomorrow Survey (TTS) of Toronto (DMG, 2011). The demand within 5 minutes were distributed randomly using a Poisson distribution. From the total demand of 5,487 trips in the loading period, we randomly extracted a percentage (e.g. 10%, 15%, 20%, 25%), as the shared vehicles demand, while the rest of the demand was assumed to travel by their own single occupancy private vehicles. In this study, we did not have any fleet size optimization and the size was set exogenously.

Although the network loading time was 15 minutes, the simulation time lasted until all passengers either arrived at their destinations or left the system. It was assumed that link-level space mean speed can be monitored, which was used by the routing agent to provide dynamic travel-time based shortest paths. The assumption is based on the fact that downtown Toronto already has enough sensors installed that can provide a quasi-real-time state of the network. We assumed that the riders left the system after their latest departure time, if they were not assigned to any vehicles. Also, as mentioned in Section 3.1, to make the ride-matching system more dynamic we assumed that the travel request time equals earliest the departure time ($t_r = e_r$).

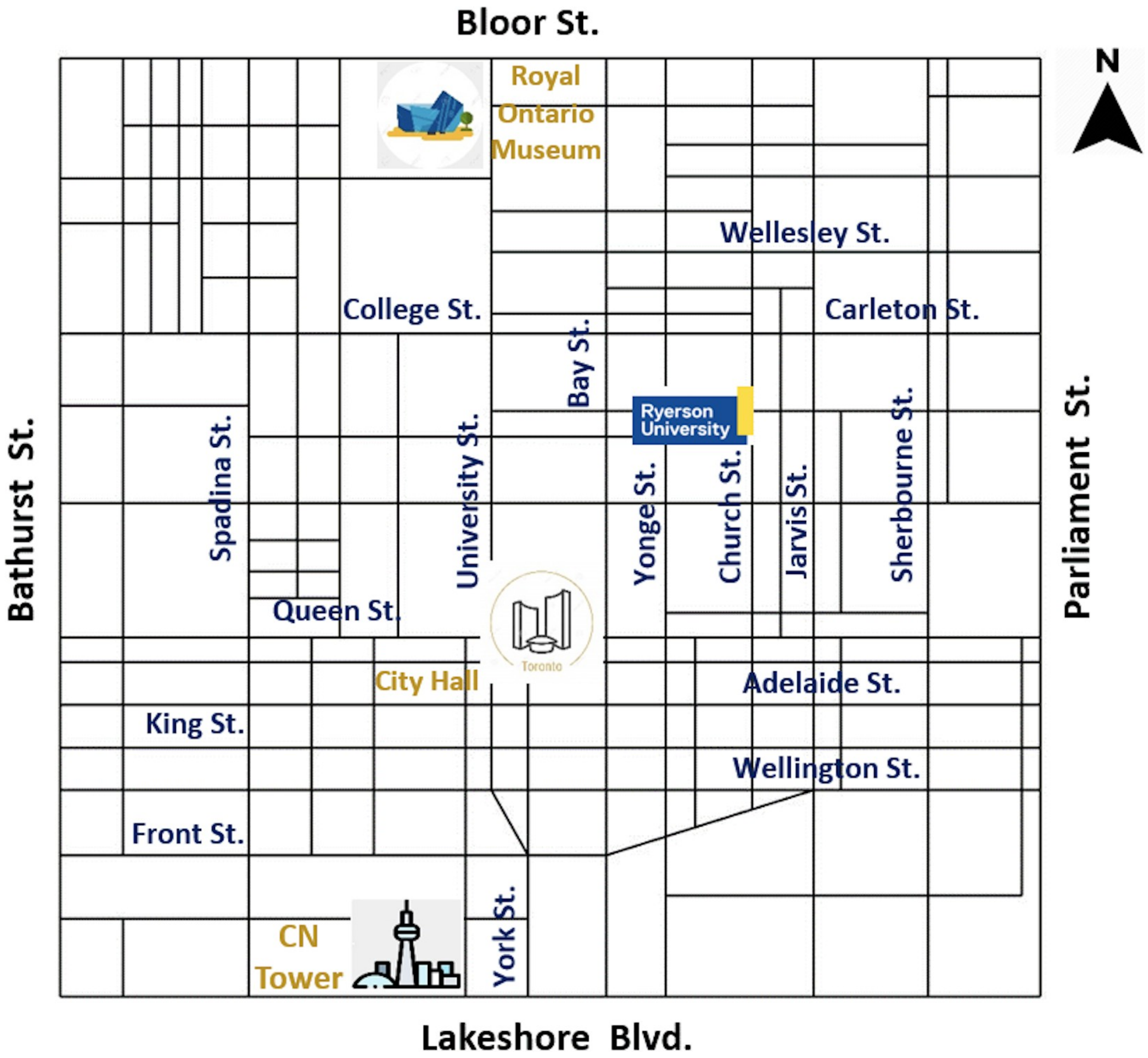

Figure 9: Downtown Toronto street network

To evaluate the performance of our GMOMatch algorithm, we compared the results with a ride-matching algorithm developed by Simonetto et al. (2019), which is an algorithm based on the linear assignment problem. We chose this algorithm because in terms of computational complexity it is one of the best algorithms in the literature, while it maintains the quality of the service. Also, unlike other recent studies that used partitioning or decomposition methods (Masoud and Jayakrishnan, 2017a; Tafreshian and Masoud, 2020), Simonetto's algorithm, like

GMOMatch, is formulated centrally and has the full view of the network. For the sake of comparison, we implemented the Simonetto's algorithm in MATLAB and applied it on the micro-traffic simulator. In both GMOMatch and Simonetto algorithms, the shared vehicles are distributed proportional to the ridesharing demand such that locations with more demand have more shared vehicles. It is worth mentioning that the Simonetto's algorithm in Simonetto et al. (2019), required rebalancing vehicles, while the GMOMatch can perform at high level without it. All simulations were implemented on three computers, including two computers with Core i7-8700 CPU, 3.20 GHz Intel with a 64-bit version of the Windows 10 operating system with 16.0 GB RAM and one computer with Core i7-6700K CPU, 4.00 GHz Intel with a 64-bit version of the Windows 10 operating system with 16.0 GB RAM.

### *4.2. Results*

The first part of this section is related to addressing the performance of the GMOMatch and providing a comparison analysis with the Simonetto's algorithm. In the second part, a detailed sensitivity analysis is conducted to show the impact of different parameters on the quality of indicators. We considered eight indicators to measure for both algorithms. Table 2 shows the indicators along with their descriptions.

Table 2: Indicators and their descriptions

| Indicators | Description |
|---|---|
| Service rate (SR) (%) | The percent of served ride requests per total requests |
| Vehicle km travelled (VKT) | Km travelled by each shared vehicle |
| Detour time (min) | The difference between shared ride travel time and direct travel time for a new request |
| Wait time (min) | The difference between passenger's pick up time and request time |
| Average traffic travel time (min) | Average all vehicles' travel time |
| Average traffic speed (km/hr) | Average all vehicles' speed |
| Average No. of assignments | Average number of assigned requests per shared vehicle over the simulation period |
| Computational time per update time (sec) | Average time it takes the algorithm is run to match a batch of ride requests with vehicles |

#### *4.2.1. GMOMatch performance and comparison analysis*

To evaluate the performance of the GMOMatch and compare it with the Simonetto, we created five scenarios by varying the fleet size (210, 230, 250, 270, 290). The shared vehicles' demand is considered to be 25% of the total demand, which came out to be 1,372 trips. The flexibility was assumed to be five minutes ($f = 5min$), vehicle capacity was four ($cap = 4$), and the update interval $\Delta = 30sec$. The simulation run-time for each scenario was between 30 and 36 hours. Fig. 10 showcases the performance of the GMOMatch and Simonetto over different parameters.

With the increase in the fleet size when the shared vehicles demand as well as the other GMOMatch parameters are fixed, due to existing more available vehicles, as expected, more ride requests can be served that leads to a higher SR. On the other hand, since both the number of vehicles and served requests increase, with the rise in the number of vehicles, no significant change is observed in other indicators, including VKT, wait time, detour time, traffic travel time, traffic speed, and No. of assignments (see Fig. 10b-10g). However, some slight fluctuations can be seen in each indicator. For example, when the fleet size is set to 230, the VKT is 7.00 km while for the size of 250 and 270, it is 6.74 and 6.91 km, respectively. One of the reasons for these slight fluctuations is that there is no vehicle relocation strategy in this ride-matching system and vehicles remain at the same location after dropping off the last passenger. Depending on new requests' origins and destinations as well as vehicles' location, vehicles may travel more or less distance to serve new requests. Thus, in some cases, despite existing more vehicles, the average No. of assignments, wait time and detour time are slightly higher. Regarding the computational time (Fig. 10h), two reasons can be pointed out. First, by increasing fleet size, the bipartite graph in step 1 as well as the vehicle graph in step 2 get larger, thus, more computations need to be done to solve the problem resulting in raising the computational time. Second, by increasing the number of vehicles, higher number of ride requests can be assigned at an update time and less unassigned requests would move to the next update time. This decreases the computational time because at each update time, the algorithm computes cost calculations for all unassigned requests. In the following, the GMOMatch and Simonetto's algorithm are compared over different indicators.

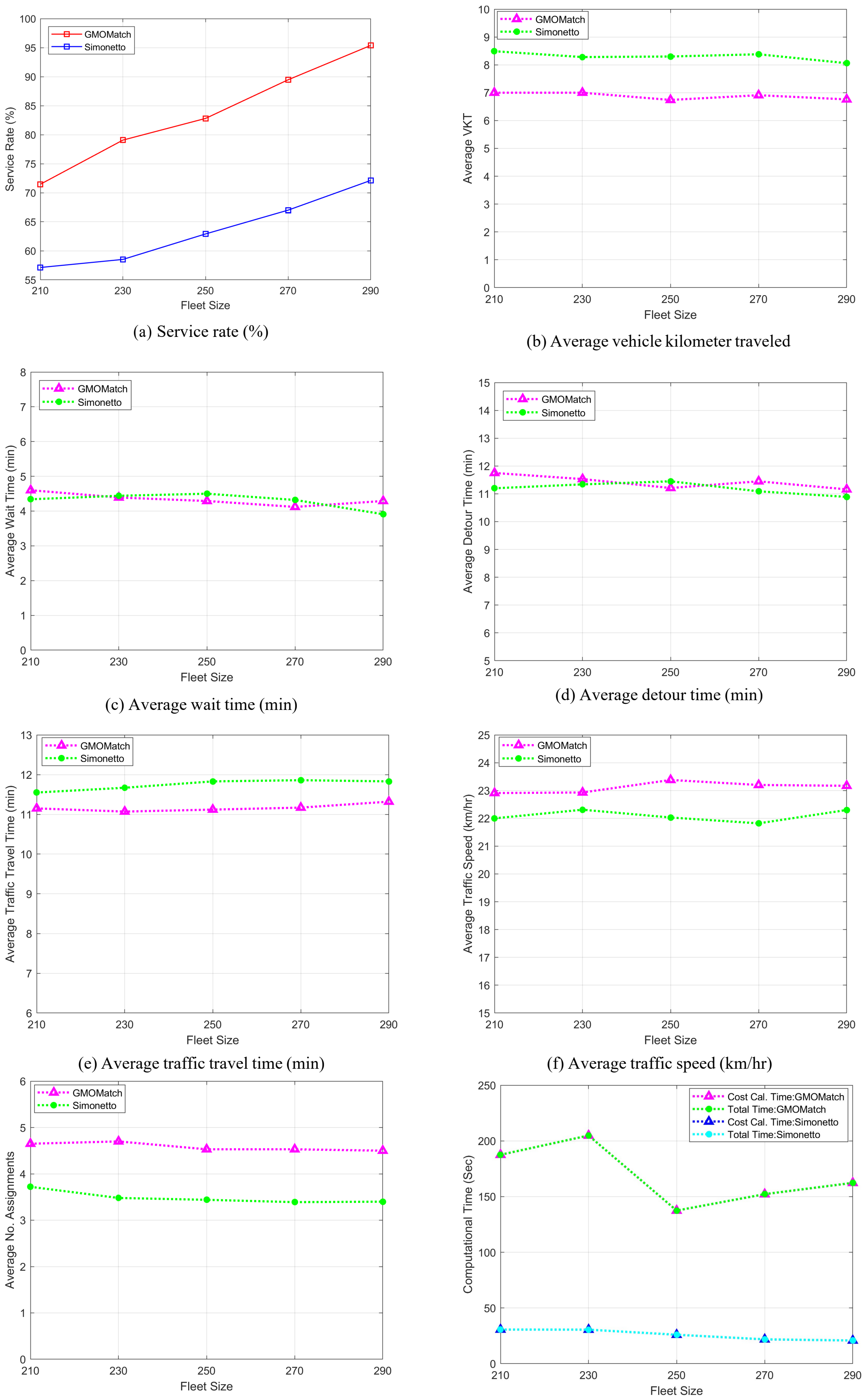

(a) Service rate (%)

(b) Average vehicle kilometer traveled

(c) Average wait time (min)

(d) Average detour time (min)

(e) Average traffic travel time (min)

(f) Average traffic speed (km/hr)

(g) Average No. assignments

(h) Computational time (sec) per call

Figure 10: Different indicators vs fleet size: demand=25%, f=5min, cap=4, $\Delta = 30sec$

Fig. 10a shows the SR for the two algorithms. As expected, by increasing fleet size when the shared vehicles demand as well as the other GMOMatch parameters are fixed, the SR improves. This is because, as a result of existing more vehicles, more ride requests can be served. For different fleet sizes, the GMOMatch yields better SR, compared to the Simonetto. For the fleet size of 290, SR for the GMOMatch was 95.39%, while for the Simonetto this number was 72.13%, which shows a 32% improvement. One of the reasons for such a large difference is that Simonetto's algorithm is based on one-to-one matching, which means that at each update time only one new request can be assigned to a vehicle despite the existence of some requests with similar itineraries at one location. Fig. 10a shows that this feature affects the SR negatively, because in congested networks the probability of having several requests with similar itineraries is high, especially with specific origins/destinations such as first-mile or last-mile problems. In contrast, the GMOMatch by combining requests with the same number of vehicles and without any rebalancing, increased the SR, indicating the high performance of the algorithm.

Fig. 10b shows the average VKT of shared vehicles over the simulation period. The GMOMatch showed lower values of VKT than the Simonetto. As an example, for the fleet size of 290, the GMOMatch showed 16.07% improvement when compared to the Simonetto, while its SR was 32% higher. One of the reasons is that in the Simonetto, because of one-to-one matching, vehicles that are enroute have to change their travel path more frequently and sometimes they may take longer paths to pick up new passengers. This repetitive change in their travel path leads to an increase in VKT. However, in the GMOMatch vehicles are usually assigned to multiple passengers instead of one. This results in having less change in the travel path and decrease in VKT. Fig. 10c and Fig. 10d demonstrate the average wait time and average detour time for different fleet sizes. Here there are no significant differences between the two algorithms. However, it is noticed that the GMOMatch with similar detour time and wait time as Simonetto served much higher number of requests.

Fig. 10e and Fig. 10f represent the average traffic travel time and the average traffic speed in the network. In Fig. 10e, average traffic travel time for the GMOMatch over different scenarios yielded better results such that for the fleet size of 290, it showed 4.26% reduction when compared to Simonetto. Also, the average traffic speed in Fig. 10f, for all scenarios in the GMOMatch showed higher speed values such that there was a 4.07% increase for the fleet size of 290. The reason is that the Simonetto's algorithm increases the number of shared vehicles in the road network because at each update time some idle shared vehicles may be assigned to new requests. These idle vehicles, which have only one occupant when they start traveling, enter the road network and worsen the traffic congestion. However, combining requests in the GMOMatch increases the vehicle occupancy rate and reduces the number of shared vehicles on the road network. This leads to improvement in the traffic travel time as well as traffic speed.

Fig. 10g shows the average No. of assignments per vehicle during the simulation period. As shown in the figure, the average No. of assignments per vehicle for the GMOMatch for different fleet size is higher than the Simonetto. As discussed earlier, combining requests in the GMOMatch in comparison with the one-to-one matching in the Simonetto improves the system's efficiency because the probability of having several requests with similar itineraries is higher in congested networks. Thus, at each update time, a vehicle in the GMOMatch algorithm may become fully-occupied, while just one request is assigned to the vehicle in the Simonetto. In the Simonetto's algorithm because of one-to-one matching, it takes more time for the vehicle to become fully-occupied.

Fig. 10h shows the average computational time per matching time for two algorithms. The computational time consists of two components: cost calculation time and total time. The cost calculation time for the GMOMatch includes creating a bipartite graph along with solving insertion method in Step 1, and creating a vehicle graph plus solving insertion method in Step 2. Total time is the summation of cost calculation time and solution time. Solution time represents the time it takes to solve one-to-one matching problem in Step 1 and maximum weight matching problem in Step 2. Most of the computational time portion is related to the cost calculation time in both algorithms. For different fleet size, the Simonetto's algorithm yielded less computational time compared to the GMOMatch. When the fleet size is 210, the GMOMatch can improve the SR by 25% at the cost of almost 5x computational time. Moreover, with the fleet size of 270 and 290, the GMOMatch can improve the SR by 33% and 32%, respectively while it scarifies the computational time by 6x and 6.8x. If the computational time is a concern for a practitioner, Simonetto's algorithm is the to go; otherwise, GMOMatch shows improvement in SR as well as other indicators and can be a good choice.

To summarize, GMOMatch reported 32% improvement when compared to the Simonetto with the fleet size of 290 and capacity of 4. Moreover, GMOMatch could reduce the VKT by 16.07% and could improve No. of assignments per vehicle by 32%. Furthermore, although the wait time and detour time were almost the same for two algorithms, GMOMatch could serve more requests compared to the Simonetto. The results also revealed that GMOMatch alleviated the traffic congestion such that the average traffic speed increased by 4.07% and the average traffic travel time reduced by 4.26%. However, the improvement of the SR was at the cost of increasing the computational time by 6.8x. There are three main reasons that can be pointed out for these improvements. The GMOMatch combines passengers and assigns them to the vehicles while the Simonetto assigns only one request to a vehicle at each matching time. This resulted in an increase in the SR. As a result of combining requests in the GMOMatch, vehicles have less enroute assignment compared to the Simonetto and also less number of vehicles are sent to the network. These reasons led to the reduction in the VKT, rise in the average No. of assignment, and improvement in the traffic congestion.

*4.2.2. Sensitivity analysis*

In the second part of the results, we conducted a detailed sensitivity analysis. Table 3 reports the results obtained by running simulations over different variables and significant algorithm parameters. Three scenarios were created by varying demand, fleet size, vehicle capacity, and flexibility to explore how changing them affect the system performance.

In the first scenario, we considered four various demand percentages, including 10%, 15%, 20%, and 25% with 150 vehicles, while keeping the other parameters constant. With the 10% demand and 150 vehicle, SR was 100% and all of the riders are served. As expected, by increasing the demand, SR reduced significantly such that with 25% demand, this number was 58.89%. It is observed that the growth in demand resulted in the increase of the VKT. Whereas, the growth had decreased the average detour time as well as wait time. In the former, although SR decreases, vehicles serve more riders during the simulation period, and thus, higher VKT was obtained. In the latter, however, with the growth of demand the probability of finding a better match increases. A better match means riders with similar itineraries are grouped and assigned to a vehicle which can lead to reduction in wait time and detour time. Finding a better match and lower values of detour time can mitigate traffic congestion because vehicles do not have to take long distances to pick up/drop off passengers. As a result, vehicles may have less change in their travel path and fewer lane changing when traveling through the road network which means lower interruption in the traffic flow and enhancement in traffic travel time and traffic speed. By increasing the demand, both cost calculation time and solution time increased. This was because the number of iterations in the GMOMatch increased, including the main iterations and iterations related to Step 2. When the demand increased, while the fleet size was fixed, the number of new requests might have exceeded the number of available vehicles. Since step 1 of the GMOMatch is one-to-one matching, the maximum number of requests that can be assigned equals the number of available vehicles. As a result of having more iterations to match the remaining requests, the cost calculation time as well as solution time increased.

In the second scenario, in Table 3, we used 25% demand, that remained fixed for all the instances. We considered three fleet sizes of 210, 230, and 250. For each fleet size, we tested three capacities of 4, 6, and 10. For each fleet size, with the increase in the capacity, SR increased. The SR was 100% for the capacity of 10, for all the fleet sizes. In a congested network with high shared vehicles demand, there might be many requests whose origins and destinations are close to each other. Therefore, they can be combined and assigned to one vehicle. This indicates the efficiency of using medium/high-capacity vehicles in the congested areas where there are enough requests with similar itineraries. Sanaullah et al. (2021) also came to a similar conclusion after evaluating the on-demand public transit in Belleville, Canada.

For each fleet size, VKT, detour time and wait time increased when the capacity increased from 4 to 6. One of the reasons is that for the capacity of 6 when the vehicles are enroute and have one or two empty spaces, they may be assigned new riders, which could lead to the rise in VKT, detour time, and wait time. For vehicles with the capacity of 10, VKT had lower values when compared to the capacity of 6, while detour time and wait time for the fleet of 230 and 250 showed slightly higher values. Traffic travel time and traffic speed for vehicles with capacity of 10, for all three fleet sizes, showed better values when compared to the capacity of 4 and 6. One of the reasons is that vehicles with capacity of 10 can transport more passengers at a time. Thus, their operational time over the simulation period might be less than vehicles with the capacity of 6 and 4. This led to a fewer number of shared vehicles on the road network, improving the average traffic travel time as well as traffic speed. Both indicators related to the computational time for all the fleet sizes, showed an increase when using vehicles with higher capacity. One of the reasons is that using vehicles with more capacity may increase the number of iterations in Step 2 of the algorithm. This is because more empty seats were available for the vehicles and more vehicle matching and consequently more requests combining may occur. However, there was a significant difference between vehicles with capacity of 10 and the other two capacity options. Such a significant difference indicates that using high-capacity vehicles despite improving the service quality is computationally expensive, especially when the shared vehicles demand is high.

Table 3: Results for different values of demand, fleet size, flexibility, and capacity

| Scenarios | Fleet Size | Demand (%) | f (min) | Δ (sec) | cap | SR (%) | VKT | Detour (min) | Waiting (min) | Traffic Travel Time (min) | Traffic Speed (km/hr) | Cost Time (s) | Solution Time (s) |
|---|---|---|---|---|---|---|---|---|---|---|---|---|---|
| 1 | 150 | 10 | 5 | 60 | 6 | 100.00 | 6.51 | 13.54 | 4.69 | 12.57 | 21.48 | 283 | 0.135 |
| | 150 | 15 | 5 | 60 | 6 | 89.16 | 6.72 | 11.90 | 4.56 | 11.96 | 22.41 | 451 | 0.214 |
| | 150 | 20 | 5 | 60 | 6 | 69.93 | 7.27 | 12.16 | 4.56 | 11.34 | 22.45 | 786 | 0.245 |
| | 150 | 25 | 5 | 60 | 6 | 58.89 | 7.46 | 11.14 | 4.21 | 10.84 | 23.20 | 963 | 0.302 |
| 2 | 210 | 25 | 5 | 60 | 4 | 79.44 | 6.70 | 10.25 | 3.61 | 11.05 | 23.18 | 538 | 0.140 |
| | 210 | 25 | 5 | 60 | 6 | 82.30 | 7.15 | 11.24 | 4.35 | 11.16 | 23.15 | 652 | 0.247 |
| | 210 | 25 | 5 | 60 | 10 | 100.00 | 6.66 | 11.47 | 4.22 | 10.79 | 23.52 | 1224 | 0.328 |
| | 230 | 25 | 5 | 60 | 4 | 85.52 | 6.54 | 10.24 | 3.52 | 11.19 | 23.05 | 523 | 0.133 |
| | 230 | 25 | 5 | 60 | 6 | 90.78 | 7.26 | 11.33 | 4.18 | 11.14 | 22.98 | 570 | 0.178 |
| | 230 | 25 | 5 | 60 | 10 | 100.00 | 6.67 | 12.02 | 4.35 | 10.82 | 23.19 | 1028 | 0.353 |
| | 250 | 25 | 5 | 60 | 4 | 91.88 | 6.47 | 10.38 | 3.47 | 11.31 | 22.67 | 416 | 0.487 |
| | 250 | 25 | 5 | 60 | 6 | 95.17 | 6.96 | 11.40 | 4.21 | 11.20 | 22.89 | 475 | 0.143 |
| | 250 | 25 | 5 | 60 | 10 | 100.00 | 6.53 | 11.68 | 4.29 | 10.92 | 23.22 | 1207 | 0.329 |
| 3 | 170 | 25 | 5 | 60 | 6 | 70.30 | 7.50 | 11.15 | 4.16 | 11.04 | 23.36 | 764 | 0.316 |
| | 170 | 25 | 10 | 60 | 6 | 76.37 | 10.32 | 14.06 | 7.08 | 10.74 | 23.05 | 1017 | 0.173 |
| | 190 | 25 | 5 | 60 | 6 | 75.05 | 7.22 | 11.35 | 4.20 | 10.95 | 22.84 | 690 | 0.286 |
| | 190 | 25 | 10 | 60 | 6 | 84.86 | 10.29 | 14.38 | 7.32 | 10.99 | 22.67 | 1127 | 0.186 |
| | 210 | 25 | 5 | 60 | 6 | 82.30 | 7.15 | 11.24 | 4.35 | 11.16 | 23.15 | 652 | 0.247 |
| | 210 | 25 | 10 | 60 | 6 | 94.22 | 10.21 | 14.09 | 6.96 | 10.96 | 23.10 | 911 | 0.234 |
| | 230 | 25 | 5 | 60 | 6 | 90.78 | 7.26 | 11.33 | 4.18 | 11.14 | 22.98 | 570 | 0.178 |
| | 230 | 25 | 10 | 60 | 6 | 100.00 | 10.25 | 13.93 | 7.19 | 11.05 | 22.65 | 910 | 0.221 |

In the third scenario, the demand was 25%, and we considered four fleet sizes, including

170, 190, 210, and 230. For each fleet size, we tested two flexibility levels (i.e., 5 and 10 minutes). For $f = 10$, the SR was higher for all fleet sizes. It is because riders keep staying in the ride-matching system 5 more minutes. During these 5 more minutes, some vehicles would become available and can be assigned to the riders. As a result of serving more riders, the VKT increased for different fleet sizes compare to $f = 5$ . The average detour time and wait time also reported higher values for $f = 10$, for all fleet sizes. This is because the riders had to wait more to be assigned and picked up. For higher average detour time, one of the reasons was that the itineraries for riders who had not been assigned are less similar to each other, which may lead to rise in the detour time. The increase in flexibility did not have any significant effect on the average traffic travel time as well as traffic speed. Both cost calculation time and solution time increased for $f = 10$. As reported for the cost calculation time, there was a significant difference between the two flexibility options. This was because by increasing the flexibility, at each update time, there are new requests and many unmatched requests from previous update times. Hence, the number of requests was higher than the number of vehicles. As discussed earlier, this increased the number of main iterations of the algorithm, which can result in an increase in the cost calculation time.

To summarize, the results revealed that in congested networks, when the demand increases, as a result of existing more requests with similar itineraries, the average wait time as well as the detour time decreases. Our findings also showed that in such high-demand networks, the use of higher capacity vehicles can result in utilizing vehicles more efficiently and decreases the fleet size. This reduction in the fleet size improved the average traffic travel time as well as the traffic speed of the network. Furthermore, increase in the passenger's flexibility led to the rise in the SR because passengers stay in the system longer than before. However, as a result of serving more requests with a fixed fleet size in this case, the average wait time as well as the detour time increased significantly.

## 5. Conclusion and future directions

We developed a novel Graph-based Many-to-One ride-Matching (GMOMatch) algorithm for solving the dynamic many-to-one ride-matching problem for shared on-demand mobility services in congested urban networks. The algorithm is iterative and has two steps. It starts by creating a bipartite graph and solves a one-to-one ride-matching problem. In the second step, a general graph, called vehicle graph, is developed and the maximum weight matching problem is solved to match the vehicles and combine the associated requests. To evaluate the performance of our algorithm, we compared it with a ride-matching algorithm developed by Simonetto et al. (2019), which is based on the linear assignment problem. We implemented two algorithms on an in-house micro-traffic simulator to compare their performance in the presence of traffic congestion. Downtown, Toronto road network was used as the case study.

The results of the study demonstrated that GMOMatch improved the SR by 32% when compared to the Simonetto's algorithm with 290 vehicles of capacity 4. Along with a higher SR, it showed either enhancement or similar performance for other indicators. VKT and No. of assignments per vehicle showed 16.07% and 32% improvement, respectively. Although there were no significant differences for the wait time and detour time between two algorithms, the GMOMatch served more requests than the Simonetto's algorithm. Comparing two algorithms also revealed that the GMOMatch alleviated the traffic congestion by increasing the average traffic speed (4.07%) and reducing the average traffic travel time (4.26%) of the traffic on the network. Overall, the GMOMatch algorithm ameliorated both service quality and traffic congestion, while its computational complexity was the same as that of the Simonetto.

To further examine the performance of GMOMatch, a detailed sensitivity analysis was performed over various parameters, including demand, fleet size, flexibility, and vehicle capacity. The sensitivity analysis showed that increasing the vehicle capacity from 4 to 10 with 210 vehicles in the congested network resulted in a 100% SR, which is a 25% improvement. Although, some slight increase in the detour time and wait time was observed, there was an improvement in other indicators, including VKT, traffic travel time and traffic speed. The use of higher capacity vehicles led to a more efficient utilization of the supply and operation of lower fleet size on the network. Our results showed that by increasing the demand and keeping the fleet size fixed, the SR reduced. However, some indicators such as detour time, traffic speed, and traffic travel time showed an improvement of 1%, 4%, and 4% , respectively. This indicates that in case of having enough demand, the operator can more likely find riders with similar origin-destinations and use the vehicle capacity more efficiently.

The proposed on-demand ride-matching algorithm has the potential to address different social, economical, and environmental aspects. It can provide high quality access to the population from different socio-economic backgrounds. Moreover, it can be operationalized for different type of urban areas, including congested as well as low-demand areas. As for the low-demand areas, different emerging shared on-demand mobility services have been introduced to enhance the efficiency as well as the attractiveness of the public transit services. For instance, the City of Belleville, Canada converted its late-night fixed-route public transit line into an on-demand transit service in September 2018 (Alsaleh and Farooq, 2020). The analysis of the service revealed that the on-demand transit service has the potential to become an effective solution for people with low accessibility in low-demand areas or during low-demand periods of the day (Sanaullah et al., 2021). GMOMatch can be useful here in ensuring efficient matching with low wait times and high system throughput. Similarly, in 2016 the Town of Innisfil, Canada concluded that operating a conventional public transit was not economically feasible. As an alternative, it started an on-demand crowdsourced transit service in 2017 that is offered by the ride-sharing app, Uber Technologies Inc. In such situation, GMOMatch has the potential to further reduce the operational cost with efficient sharing of trips, while maintaining low wait and travel times. The shared ride-matching service is more economical for users because it is less expensive than single passenger ridehailing trips or purchasing and maintaining a vehicle. Moreover, the fare is split among users that makes it more affordable, while it remains an acceptable level of convenience (Machado et al., 2018). As a result of offering high-quality service with reasonable price, more people are encouraged to use such on-demand shared services. Recent studies have shown that the shared on demand mobility services have the potential to reduce the emissions by 30% after replacing personal vehicles (Buerstlein et al., 2021). Considering the appropriate performance of the proposed algorithm in reducing VKT and improving traffic congestion, the proposed on-demand ride-matching service can provide cities with more sustainable and environment-friendly travel mode.

We expect that this work will be useful for the design and operations of on demand shared mobility services using medium/high capacity vehicles in dense urban areas, where congestion is a recurrent issue. Moreover, the proposed algorithm could be useful for the first mile/last mile service for freight or passenger systems, where an origin or destination is fixed.

In less congested areas, combining the requests may increase the VKT, wait time and detour time. The probability of finding requests with similar itineraries is lower, either when the rate of the shared vehicles demand is low or when the update time interval to run the matching algorithm is very short. One of the ways to address these limitations is to develop a reinforcement learning system that can dynamically assess the need of Step 2 in GMOMatch, based on the changing spatiotemporal dynamics of demand. This way, only the one-to-one matching problem in Step 1 is run iteratively and the operator can consistently keep the service quality at high level.

All of the cost computations in this study have been calculated centrally. In the future, distributed computing and different graph partitioning methods can be used to reduce the cost calculation time. A promising option to use here is the distributed system designed by Farooq and Djavadian (2019). Such distributed systems can significantly decrease the computational time by only performing local level computations. As mentioned, our data in this study were generated from a simulation. Having access to the real data can give better insights into user behaviour, which would be helpful for adjusting the algorithm parameters to improve the efficiency of the ride-matching system. Combining the two steps of the proposed algorithm can be explored using the k-partite matching approaches suggested by Anderson and Farooq (2017). The proposed ride-matching algorithm in this study does not have any vehicle relocation, inclusion of which can further improve the performance. Predicting demand and developing a vehicle relocation system using machine learning techniques can significantly enhance the service quality.

**Acknowledgement**

This research is funded by the Canadian Urban Transit Research and Innovation Consortium (CUTRIC) and Ryerson University. We would like to thank Dr. Shadi Djavadian for the initial implementation of the base traffic micro-simulator adapted in this study and her continued support in the implementation of the proposed algorithms in the simulator.

## References

N. Agatz, A. L. Erera, M. W. Savelsbergh, and X. Wang. Dynamic ride-sharing: A simulation study in metro atlanta. *Procedia-Social and Behavioral Sciences*, 17:532–550, 2011.

N. Agatz, A. Erera, M. Savelsbergh, and X. Wang. Optimization for dynamic ride-sharing: A review. *European Journal of Operational Research*, 223(2):295–303, 2012.

J. Alonso-Mora, S. Samaranayake, A. Wallar, E. Frazzoli, and D. Rus. On-demand high-capacity ride-sharing via dynamic trip-vehicle assignment. *Proceedings of the National Academy of Sciences*, 114(3):462–467, 2017.

N. Alsaleh and B. Farooq. Machine learning based demand modelling for on-demand transit services: A case study of belleville, ontario. *arXiv e-prints*, pages arXiv–2010, 2020.

P. Anderson and B. Farooq. A generalized partite-graph method for transportation data association. *Transportation Research Part C: Emerging Technologies*, 76:150–169, 2017.

M. Baza, M. Nabil, M. Ismail, M. Mahmoud, E. Serpedin, and M. A. Rahman. Blockchain-based charging coordination mechanism for smart grid energy storage units. In *2019 IEEE International Conference on Blockchain (Blockchain)*, pages 504–509. IEEE, 2019.

D. Bertsimas, P. Jaillet, and S. Martin. Online vehicle routing: The edge of optimization in large-scale applications. *Operations Research*, 67(1):143–162, 2019.

J. Buerstlein, D. Lopez, and B. Farooq. Exploring first-mile on-demand transit solutions for north american suburbia: A case study of markham, canada. *Transportation Research Part A: Policy and Practice*, pages 1–29, 2021.

J. Castiglione, D. Cooper, B. Sana, D. Tischler, T. Chang, G. D. Erhardt, S. Roy, M. Chen, and A. Mucci. Tncs & congestion. 2018.

A. Chehri and H. T. Mouftah. Autonomous vehicles in the sustainable cities, the beginning of a green adventure. *Sustainable Cities and Society*, 51:101751, 2019.

F. Dandl, B. Bracher, and K. Bogenberger. Microsimulation of an autonomous taxi-system in munich. In *2017 5th IEEE International Conference on Models and Technologies for Intelligent Transportation Systems (MT-ITS)*, pages 833–838. IEEE, 2017.

Z. De Giovanni, Di Summa. Methods and Models for Combinatorial Optimization, 2020. URL https://www.math.unipd.it/~luigi/courses/metmodoc1920/m07.assTum.en.pdf/.

S. Djavadian and B. Farooq. Distributed dynamic routing using network of intelligent intersections. In *ITS Canada Annual General Meeting Conference, Niagara Falls*, 2018.

DMG. Transportation Tomorrow Survey, 2011. URL http://www.dmg.utoronto.ca/transportationtomorrowsurvey/.

B. Farooq and S. Djavadian. Distributed traffic management system with dynamic end-to-end routing, 2019. URL https://patentscope.wipo.int/search/en/detail.jsf?docId=WO2020257926. U.S. provisional patent 62/865,725.

G. Feng, G. Kong, and Z. Wang. We are on the way: Analysis of on-demand ride-hailing systems. *Available at SSRN 2960991*, 2017.

M. Guériau, F. Cugurullo, R. Acheampong, and I. Dusparic. Shared autonomous mobility-on-demand: Learning-based approach and its performance in the presence of traffic congestion. *IEEE Intelligent Transportation Systems Magazine*, 2020.

L. Häme. An adaptive insertion algorithm for the single-vehicle dial-a-ride problem with narrow time windows. *European Journal of Operational Research*, 209(1):11–22, 2011.

S. C. Ho, W. Y. Szeto, Y.-H. Kuo, J. M. Leung, M. Petering, and T. W. Tou. A survey of dial-a-ride problems: Literature review and recent developments. *Transportation Research Part B: Methodological*, 111:395–421, 2018.

Y. Huang, K. M. Kockelman, V. Garikapati, L. Zhu, and S. Young. Use of shared automated vehicles for first-mile last-mile service: Micro-simulation of rail-transit connections in austin, texas. *Transportation Research Record*, page 0361198120962491, 2020.

J. Jung, R. Jayakrishnan, and J. Y. Park. Dynamic shared-taxi dispatch algorithm with hybrid-simulated annealing. *Computer-Aided Civil and Infrastructure Engineering*, 31(4):275–291, 2016.

S. Kudva, R. Norderhaug, S. Badsha, S. Sengupta, and A. Kayes. Pebers: Practical ethereum blockchain based efficient ride hailing service. In *2020 IEEE International Conference on Informatics, IoT, and Enabling Technologies (ICIoT)*, pages 422–428. IEEE, 2020.

M. Li, J. Weng, A. Yang, W. Lu, Y. Zhang, L. Hou, J.-N. Liu, Y. Xiang, and R. H. Deng. Crowdbc: A blockchain-based decentralized framework for crowdsourcing. *IEEE Transactions on Parallel and Distributed Systems*, 30 (6):1251–1266, 2018.

M. Liu, Z. Luo, and A. Lim. A branch-and-cut algorithm for a realistic dial-a-ride problem. *Transportation Research Part B: Methodological*, 81:267–288, 2015.

M. Lokhandwala and H. Cai. Dynamic ride sharing using traditional taxis and shared autonomous taxis: A case study of nyc. *Transportation Research Part C: Emerging Technologies*, 97:45–60, 2018.

G. Lyu, W. C. Cheung, C.-P. Teo, and H. Wang. Multi-objective online ride-matching. *Available at SSRN 3356823*, 2019.

C. A. S. Machado, N. P. M. de Salles Hue, F. T. Berssaneti, and J. A. Quintanilha. An overview of shared mobility. *Sustainability*, 10(12):4342, 2018.

N. Masoud and R. Jayakrishnan. A decomposition algorithm to solve the multi-hop peer-to-peer ride-matching problem. *Transportation Research Part B: Methodological*, 99:1–29, 2017a.

N. Masoud and R. Jayakrishnan. A real-time algorithm to solve the peer-to-peer ride-matching problem in a flexible ridesharing system. *Transportation Research Part B: Methodological*, 106:218–236, 2017b.

A. Mourad, J. Puchinger, and C. Chu. A survey of models and algorithms for optimizing shared mobility. *Transportation Research Part B: Methodological*, 2019.

A. Najmi, D. Rey, and T. H. Rashidi. Novel dynamic formulations for real-time ride-sharing systems. *Transportation research part E: logistics and transportation review*, 108:122–140, 2017.

M. Nourinejad and M. J. Roorda. Agent based model for dynamic ridesharing. *Transportation Research Part C: Emerging Technologies*, 64:117–132, 2016.

S. Oh, R. Seshadri, D.-T. Le, P. C. Zegras, and M. E. Ben-Akiva. Evaluating automated demand responsive transit using microsimulation. *IEEE Access*, 8:82551–82561, 2020.

E. Özkan and A. R. Ward. Dynamic matching for real-time ride sharing. *Stochastic Systems*, 10(1):29–70, 2020.

X. Qian, W. Zhang, S. V. Ukkusuri, and C. Yang. Optimal assignment and incentive design in the taxi group ride problem. *Transportation Research Part B: Methodological*, 103:208–226, 2017.

X. Qian, T. Lei, J. Xue, Z. Lei, and S. V. Ukkusuri. Impact of transportation network companies on urban congestion: Evidence from large-scale trajectory data. *Sustainable Cities and Society*, 55:102053, 2020.

H. Ritchie. Urbanization. *Our World in Data*, 2018. https://ourworldindata.org/urbanization.

S. Sachan, S. Deb, and S. N. Singh. Different charging infrastructures along with smart charging strategies for electric vehicles. *Sustainable Cities and Society*, 60:102238, 2020.

I. e. c. Sanaullah, N. e. c. Alsaleh, S. Djavadian, and B. Farooq. Spatio-temporal analysis of on demand transit: A case study of belleville, canada. *Transportation Research Part A: Policy and Practice*, 145:284–301, 2021.

D. Sánchez, S. Martínez, and J. Domingo-Ferrer. Co-utile p2p ridesharing via decentralization and reputation management. *Transportation Research Part C: Emerging Technologies*, 73:147–166, 2016.

Saunders. Weighted maximum matching in general graphs - file exchange - matlab central. https://www.mathworks.com/matlabcentral/fileexchange/ 42827-weighted-maximum-matching-in-general-graphs, 2013. (Accessed on 12/03/2020).

M. W. Savelsbergh and M. Sol. The general pickup and delivery problem. *Transportation science*, 29(1):17–29, 1995.

S. Shaheen, A. Cohen, M. Randolph, E. Farrar, R. Davis, and A. Nichols. Shared mobility policy playbook. 2019.

A. Simonetto, J. Monteil, and C. Gambella. Real-time city-scale ridesharing via linear assignment problems. *Transportation Research Part C: Emerging Technologies*, 101:208–232, 2019.

A. Tafreshian and N. Masoud. Trip-based graph partitioning in dynamic ridesharing. *Transportation Research Part C: Emerging Technologies*, 114:532–553, 2020.

A. Tafreshian, N. Masoud, and Y. Yin. Frontiers in service science: Ride matching for peer-to-peer ride sharing: A review and future directions. *Service Science*, 12(2-3):44–60, 2020.

H. Wang and H. Yang. Ridesourcing systems: A framework and review. *Transportation Research Part B: Methodological*, 129:122–155, 2019.

H. Yu, X. Jia, H. Zhang, X. Yu, and J. Shu. Psride: Privacy-preserving shared ride matching for online ride hailing systems. *IEEE Transactions on Dependable and Secure Computing*, 2019.